\documentclass[10pt,aps,pra,twocolumn,groupedaddress,reprint,amsmath,amssymb,showkeys,floatfix]{revtex4-2}

\usepackage{graphicx}
\usepackage{siunitx}
\usepackage{booktabs}
\usepackage{pgfplots} 
\usepackage{tikz}
\usetikzlibrary{decorations.pathmorphing,decorations.markings,arrows.meta}
\usepackage[version=4]{mhchem}
\usepackage[colorlinks=true,linkcolor=blue, citecolor=blue, urlcolor=blue, bookmarks]{hyperref}
\usepackage[dvipsnames]{xcolor}
\usepackage{comment}
\usepackage{tikz}
\usepackage{xcolor}
\definecolor{beaublue}{rgb}{0.7, 0.75, 0.71}   
\definecolor{bubblegum}{rgb}{1.0, 0.5, 0.31}   

\definecolor{beaublueBlack70}{rgb}{0.21, 0.225, 0.213}   
\definecolor{beaublueTint15}{rgb}{0.955, 0.9625, 0.9575} 
\definecolor{darkgreen}{RGB}{0,80,45}
\usepackage{multirow}

\DeclareMathOperator{\sgn}{sgn}
\newcommand{\RE}{\mathrm{Re}}
\newcommand{\IM}{\mathrm{Im}}
\newcommand{\til}{~}
\newcommand{\eF}{{E^0_\mathrm{F}}}
\newcommand{\eFs}{{E^0_{\mathrm{F} \sigma}}}
\newcommand{\gtwoD}{{ \ln\left(1/k_\mathrm{F}^0a_\mathrm{2D}\right)}}
\newcommand{\gthreeD}{{1/{k_\mathrm{F}^0a_\mathrm{F}}}}
\newcommand{\kF}{{k_\mathrm{F}^0}}
\newcommand{\kFs}{{ k^0_{\mathrm{F}\sigma} }}
\newcommand{\be}{{\varepsilon_0}}

\newcommand{\vQ}{{\mathbf{Q}}}
\newcommand{\vq}{{\mathbf{q}}}
\newcommand{\vk}{{\mathbf{k}}}

\newcommand{\omb}{{\bar{\omega}}}
\newcommand{\Omb}{{\bar{\Omega}}}
\newcommand{\om}{{\omega}}
\newcommand{\Om}{{\Omega}}
\newcommand{\omt}{{\tilde{\omega}}}
\newcommand{\Qfflo}{{Q^*}}
\newcommand{\sQfflo}{{Q_\mathrm{FF} }}
\newcommand{\sQfflosq}{{Q^2_\mathrm{FF} }}
\newcommand{\vff}{{ v_\mathrm{FF} }}

\newcommand{\vQfflo}{{\mathbf{Q}_\mathrm{FF} }}
\newcommand{\muL}{{ \tilde{\mu}_{\sigma} }}
\newcommand{\vL}{{ {v}_{\mathrm{F}\sigma} }}

\newcommand{\vvL}{{ \mathbf{{v}}_{\mathrm{F}\sigma} }}
\newcommand{\vvLup}{{ \mathbf{{v}}_{\mathrm{F}\uparrow} }}
\newcommand{\vvLdn}{{ \mathbf{{v}}_{\mathrm{F}\downarrow} }}
\newcommand{\vLup}{{ {{v}}_{\mathrm{F}\uparrow} }}
\newcommand{\vLdn}{{ {{v}}_{\mathrm{F}\downarrow} }}

\newcommand{\vLupx}{{ {v}^{(x)}_{\mathrm{F}\uparrow} }}
\newcommand{\vLupy}{{ {v}^{(y)}_{\mathrm{F}\uparrow} }}
\newcommand{\kL}{{ {k}_{\mathrm{F}\sigma} }}
\newcommand{\kLsq}{{ {k}^2_{\mathrm{F}\sigma} }}
\newcommand{\kLb}{{ {k}_{\mathrm{F}\bar{\sigma}} }}
\newcommand{\vkL}{{ {\mathbf{k}}_{\mathrm{F}\sigma} }}

\newcommand{\kLup}{{ {k}_{\mathrm{F}\uparrow} }}
\newcommand{\kLupsq}{{ {k}^2_{\mathrm{F}\uparrow} }}
\newcommand{\kLdn}{{ {k}_{\mathrm{F}\downarrow} }}
\newcommand{\kLdnsq}{{ {k}^2_{\mathrm{F}\downarrow} }}
\newcommand{\vkLup}{{ \mathbf{{k}}_{\mathrm{F}\uparrow} }}
\newcommand{\vkLdn}{{ \mathbf{{k}}_{\mathrm{F}\downarrow} }}

\newcommand{\kdn}{{k_{\downarrow}}}

\newcommand{\Gt}{{\tilde{G}}}
\newcommand{\xit}{{\tilde{\xi}}}
\newcommand{\mut}{{\tilde{\mu}}}

\begin{document}

\preprint{APS/123-QED}

\title{Fulde-Ferrell-Larkin-Ovchinnikov transition and quantum criticality in polarized Fermi gases}

\author{Francesco Pirolo}
\email{francesco.pirolo2@unibo.it}
\author{Leonardo Pisani}
\email{leonardo.pisani2@unibo.it}
\author{Pierbiagio Pieri}
\email{pierbiagio.pieri@unibo.it}

\affiliation{Dipartimento di Fisica e Astronomia, Università di Bologna, via Irnerio 46, I-40126 Bologna, Italy}
\affiliation{INFN, Sezione di Bologna, Viale Berti Pichat 6/2, I-40127, Bologna, Italy}

\date{\today}
\begin{abstract}
We investigate the zero-temperature transition from the polarized normal phase to the FFLO state in a two-dimensional Fermi gas by means of a diagrammatic $t$-matrix approach. We first show that the standard non-self-consistent theory produces an unphysical  critical polarization because of a severe violation of the Luttinger theorem. Motivated by this observation, we introduce a minimal self-consistent extension that largely restores compliance with the Luttinger theorem while preserving the analytical simplicity of the original formalism. This leads to a physically consistent  critical polarization over the whole interaction range.
Building on this improved description, we characterize the quantum critical behavior of the FFLO transition through the quasiparticle decay rates, quasiparticle weights, momentum distributions, and the critical dynamics of both fermionic and bosonic degrees of freedom. We further compare the two- and three-dimensional systems, showing that their critical properties can be understood within a unified geometrical picture based on the nesting of the majority and minority Fermi surfaces. 
Finally, for a three-dimensional lattice, we demonstrate within the Hertz-Millis framework that vertex corrections are irrelevant,
thereby placing the FFLO quantum phase transition in the mean-field universality class, in close analogy with the itinerant antiferromagnet transition. 
\end{abstract}

\maketitle

\section{Introduction}
\label{sec:INTRO}

The advent of high-$T_c$ superconductors and the observation of a new metallic phase called ``strange metal" has spurred both experimental and theoretical interest in metallic matter with properties deviating from Fermi liquid theory.
In itinerant systems non-Fermi liquid behavior generally takes place in the proximity of a quantum critical point
\cite{Varma2002,Vojta2007,Senthil2008,Senthil2015,BK-2016}. 

The original approach by Hertz  \cite{Hertz-1976} on metallic quantum  criticality has revealed a number of shortcomings, chief of all the assumption of a well-behaved expansion of the  Landau-Ginzburg-Wilson (LGW) free energy in the presence
of gapless fermionic excitations \cite{Chubukov-2003}. Whilst Hertz's theory predictions have proven to be appropriate in most three-dimensional (3D) itinerant phase transitions, it is now known that the generalization of the LGW approach to the pure quantum regime breaks down in two-dimensions \cite{BKV-2005,Chubukov-2003,Sachdev2011}. A number of works have reexamined the problem of the instabilities of two-dimensional (2D) metals towards ferromagnetism \cite{BKV1997,Chubukov-2004-FM}, anti-ferromagnetism \cite{Chubukov-2004-AFM,Trembley2012,Sachdev2012}, charge density waves \cite{Holder2014} and Ising-nematic ordering \cite{Holder-2015}.

The instability towards the superconducting state can also display critical non-Fermi liquid behavior when a magnetic field is present. 
In solid-state superconductors (and disregarding orbital effects), a magnetic field Zeeman-splits the two spin states, producing different occupations and eventually leading to the Pauli paramagnetic limitation of conventional superconductivity, also known as Clogston-Chandrasekhar limit \cite{Chandrasekhar1962,Clogston1962}. In ultracold Fermi gases the protocol is reversed: one can prepare unequal populations of two selected hyperfine states from the outset, thereby realizing a polarized Fermi gas \cite{Zwierlein2006,Partridge2006,Shin2006}. The imbalance produces a mismatch between the two Fermi surfaces (FS's) and favors unconventional finite-momentum pairing. 
The latter mechanism, now known as FFLO pairing, was proposed independently by Fulde and Ferrell and by Larkin and Ovchinnikov \cite{Fulde1964,Larkin1964} (see \cite{Casalbuoni2004,Matsuda2007,Combescot2007} and \cite{Kinnunen2018,Strinati2018} for recent reviews). In the ordered phase, finite-momentum pairing produces a spatially modulated order parameter, with the simplest Fulde-Ferrell and Larkin-Ovchinnikov forms given respectively by
$\Delta(\mathbf r)=\Delta_0 e^{i\mathbf q\cdot\mathbf r}$ and $\Delta(\mathbf r)=\Delta_0\cos(\mathbf q\cdot\mathbf r)$.
At the transition point the FF and LO channels are degenerate, and we shall therefore refer to the finite-momentum pairing instability as FFLO. Above the transition point and depending on coupling strength and polarization, the systems remains in a normal polarized Fermi liquid state. 

The search for FFLO order has remained active in solid-state systems, including heavy-fermion compounds, organic superconductors, and iron-based superconductors \cite{Bianchi2003,Kenzelmann2008,Lortz2007,Wright2011,Mayaffre2014,Koutroulakis2016,Cho2017,Kasahara2020}. In ultracold gases, the closest experimental realization has been achieved in quasi-one-dimensional polarized Fermi gases, while recent advances in quantum gas microscopy and low-dimensional platforms provide new tools for detecting pairing correlations \cite{Liao2010,Olsen2015,Revelle2016,2015-Thomas,Bakr-2016,Sundar2020,Yao-2025}.

At finite temperatures, thermal fluctuations are known to destabilize the FFLO transition in continuum systems with rotational invariance, in both two and three dimensions \cite{Radzihovsky-2009,Radzihovsky-2011} (see also \cite{Jakubczyk-2017}). 
This instability originates from the spherical degeneracy of the ordering wave vector $\vq$, which gives rise to exceptionally soft Goldstone modes that strongly enhance fluctuation effects
\cite{Radzihovsky-2011} (see also \cite{1975-Brazovskii,Shimahara-1998,Ohashi-2002}).
At $T=0$, however, the additional effective dimension associated with quantum dynamics suppresses these fluctuations, thereby restoring the stability of the FFLO phase in both 2D and 3D \cite{Radzihovsky-2011}.

We therefore focus on the zero temperature limit and approach the instability from the normal side, where signatures of finite-momentum pairing can be identified as an enhancement of the pair susceptibility at a nonzero wave vector $Q_{\mathrm{FF}}$ \cite{Pini2021}, down to the transition point where this enhancement becomes a divergence.

At the quantum critical point both collective (bosonic) and single-particle (fermionic) excitations becomes critically soft and a non-Fermi liquid behavior ensues \cite{BKV-2005}.
The concomitance of gapless bosonic and fermionic excitations is responsible for the breakdown of the core assumption of Fermi liquid theory, that is the one-to-one correspondence between the one-particle states of the interacting and non-interacting gas \cite{Varma2002}.
As a consequence one expects to see non-Fermi liquid features reflected in single-particle observables like momentum distributions, quasi-particle weights and decay rates, stemming from non-analytical behaviors of fermionic and bosonic self-energies in the low-frequency limit  \cite{Senthil2008}.

Previous work on the zero-temperature polarized Fermi gases has encompassed a number of theoretical approaches, from mean-field theory \cite{Chandrasekhar1962,Clogston1962,Sarma1963,SheehyRadzihovsky2006,Sheehy2007,Parish2007,ChevyMora2010,RadzihovskySheehy2010} and dynamical mean-field theory \cite{Kinnunen2018} to Quantum Monte Carlo \cite{Lobo2006,Pilati2008,Troyer2016,Vitali2022} and diagrammatic approaches \cite{Levin-2007,Takashi2012,Urban2014,Tajima2014,Zwerger2018,Pini2021}.
In this work we are concerned with the latter methodology in the form of the $t$-matrix approximation, that has been shown to provide a satisfactory description of the BCS-BEC crossover \cite{Strinati2018} both in its non-self-consistent \cite{Perali2002,Pieri2004} and self-consistent versions \cite{Pini2019}. 

Specifically, we are interested in extending the recent $t$-matrix study of the 3D polarized Fermi gas at $T=0$ \cite{Pini2023} to two dimensions.
That work was primarily concerned with the computation of the phase diagram {\it critical polarization versus coupling}, separating the normal from the condensed phase, and with the breakdown of Fermi liquid theory at the FFLO instability
through a detailed analysis of the quasi-particle weight and decay rate on the critical line.
In addition, the shortcomings of the non-self-consistent approach in the presence of a finite polarization were pointed out.
The main origin of these shortcomings  was attributed to the violation of the Luttinger theorem, a crucial requisite for a physically meaningful description of polarized Fermi gases \cite{Urban2014,Pantel2014}.

Within mean-field theory, the zero-temperature phase diagram of a polarized two-dimensional Fermi gas across the BCS–BEC crossover was investigated in Ref.~\cite{2008-Conduit}. As in the three-dimensional case \cite{Sheehy2007}, the FFLO phase was found to occupy only a relatively narrow region of the polarization–coupling plane, lying between the normal phase and a phase-separation region.
Furthermore, analytical expressions for the static pairing susceptibility, the FFLO ordering wave vector, and the critical pseudo-magnetic field at the mean-field normal-FFLO transition, were derived in Ref.~\cite{Sheehy2015}. 

In Ref.~\cite{Samokhin2006} quantum critical fluctuations were taken into account at the level of the non-self-consistent $t$-matrix approach and their effect on quasi-particle decay rates examined. However, the low-energy form of the fluctuation propagator was  assumed to possess the same Landau damping term as in the 3D setting. Moreover, only the non-isotropic case of multiple but isolated ordering wave-vectors (owing to the discrete geometry) was considered.
More recently, the authors of Ref.~\cite{Piazza2016} re-assessed the problem
of the quasi-particle decay rate in 2D
within the non-self-consistent $t$-matrix approach and found a non-Fermi liquid behavior of the fermionic self-energy at low frequency with fractional exponent $2/3$, assuming a small polarization and weak coupling strength.
Subsequent work in Ref.~\cite{Pimenov_2018} adopted a renormalization group analysis at the one-loop order, confirming the fermionic single-particle results and the scaling of the bosonic propagator of Ref.~\cite{Piazza2016}.

Purpose of the present work is to examine the zero-temperature polarized Fermi gas in two-dimensions across all coupling strengths and polarization values within the normal phase and down to the FFLO instability. 
At the quantum critical point we carefully analyze the non-Fermi liquid behavior manifested in the quasi-particle weight, decay rate and momentum distribution. We therefore extend the work in \cite{Pini2023} to two dimensions by adopting a minimally self-consistent $t$-matrix approximation that essentially (albeit not exactly) guarantees compliance with the Luttinger theorem.  The inclusion of such minimal degree of self-consistency will allow us to go beyond the analysis of Ref.~\cite{Piazza2016} by relaxing the small polarization and weak-coupling assumptions. 

Implementation of full self-consistency
in a 2D setting proves to be particularly challenging owing to the slow decay of frequency integrands in the ultraviolet as well as to the well-known enhanced critical fluctuations in the infrared.
Therefore, we  employ a simplified scheme which still guarantees the following fundamental features: 
(i) a reliable description of the interaction in the normal state, (ii) substantial consistence with the Luttinger theorem, (iii) a physically sound critical polarization, (iv)  feasibility of analytical calculations.
The minimal self-consistent $t$-matrix approach (MSCT) here adopted satisfies  the above requirements.
The downside of not implementing the full self-consistency is reflected in a critical polarization that does not deviate significantly from that obtained within mean-field theory. We will explain in detail the origin of this drawback later on.

Our main results are as follows: (i) the computation of the {\it critical polarization versus coupling} from the polaronic limit to the superfluid transition, (ii) the numerical as well  as analytical evaluation of the fermionic self-energies at momenta on and off the FS (associated with non-Fermi-liquid and marginal-Fermi-liquid behaviors respectively, (iii) the evolution of the momentum distributions, quasi-particle decay rates  and universally scaling quasi-particle weights down to the critical line, and finally (iv) the identification of  the bosonic (order parameter) and fermionic dynamical exponents. In this respect, we emphasize that, as far as the calculation of critical behavior is concerned, the present MSCT approach corresponds to what in the renormalization-group framework is known as the RPA or one-loop level \cite{Holder-2015-2}. Accordingly, the critical exponents reported below should be understood as predictions within this level of approximation, while a complete characterization of the asymptotic critical behavior would require a renormalization-group analysis beyond it. On the other hand, the microscopic approach provides access to system-specific, non-universal quantities, such as the critical polarization and the momentum distributions. Independently of this, determining the critical properties predicted by the approximate MSCT theory remains important both for a complete characterization of the theory itself and for providing an analytical benchmark against which numerical solutions of the MSCT equations can be tested as the critical point is approached.

Finally, on a 3D lattice, motivated by the analogy existing in this case between the FF state and the itinerant antiferromagnet, we formulate Hertz's theory in the finite-momentum pairing channel to demonstrate that all vertex corrections above the bare bosonic propagator (mode-mode coupling terms in \cite{Hertz-1976}) are irrelevant in the renormalization group flow, thereby placing the FF transition in a 3D lattice  in the same mean-field universality class as the itinerant antiferromagnet.

The paper is organized as follows. In Sec. II we introduce the model, the $t$-matrix formalism, the non-self-consistent and minimal self-consistent schemes, and the generalized Thouless criterion used to locate the FFLO instability. In Sec. III we describe the numerical implementation of the self-energy and the momentum-distribution calculations. In Sec. IV we present the two-dimensional critical polarization and compare it with the 3D counterpart, emphasizing the role of Luttinger-theorem violations and their partial correction within the minimal self-consistent scheme. In Sec. V we study the signatures of non-Fermi-liquid behavior in quasiparticle decay rates, quasiparticle weights, and momentum distributions. In Sec. VI we analyze the critical dynamical exponents of the bosonic and fermionic sectors both in 2D and in 3D, and evaluate vertex corrections on top of the bare bosonic propagator within Hertz's theory for the 3D lattice case. In addition, we provide a schematic comparison of the FFLO transition with other metallic quantum critical systems.
In Section VII we draw our conclusions.

The appendices contain more technical material.
In appendix \ref{sec:Thoulcrit} we provide an analytical expression for the fluctuation propagator in the complex frequency space  as well as in the static limit and extract the critical exponent for the correlation length. 
Appendix \ref{sec:ImGammaRet} provides a detailed analysis  of the pair spectral weight function. 
Appendix \ref{sec:gammalowE} obtains the critical pair fluctation propagator for small frequencies and momenta.
In appendix \ref{app:imsigretonoffFS}
the imaginary part of the retarded self-energy is evaluated analytically for small frequencies and momenta on and off the FS of the minority species.
In appendix \ref{app:imsigret3D} we derive an analytical expression for the imaginary part of the retarded self-energy at the 3D FFLO transition for general momenta and small frequencies.
In appendix \ref{app:4bosvert} we compute the lowest order vertex correction to the fluctuation propagator of the 3D FFLO transition and show its irrelevance in the renormalization group flow associated  with the Hertz-Millis effective action in the pairing channel.
In appendix \ref{sec:wclim} an analytical expression is obtained for the critical polarization in the weak-coupling limit of the non-self-consistent $t$-matrix approach.

\section{Theoretical Formalism}
\label{sec:Formalism}

\noindent

We consider a homogeneous two-dimensional (2D) system of spin-1/2 fermions of mass $m$ interacting through an attractive contact interaction at zero temperature. This system is  described by the following Hamiltonian (in the following the reduced
Planck constant $\hbar$ is set equal to unity)
\begin{equation}
\begin{aligned}
&\hat{H}= \sum_{\sigma} \int d\mathbf{r}\, 
\hat{\psi}^\dagger_{\sigma}(\mathbf{r}) 
\left(-\frac{\nabla^2}{2m}\right) 
\hat{\psi}_{\sigma}(\mathbf{r})
\\&+ v_0 \int d\mathbf{r}\,
\hat{\psi}^\dagger_{\uparrow}(\mathbf{r})
\hat{\psi}^\dagger_{\downarrow}(\mathbf{r})
\hat{\psi}_{\downarrow}(\mathbf{r})
\hat{\psi}_{\uparrow}(\mathbf{r}),
\end{aligned}
\label{eq:Hamiltonian}
\end{equation} 
where $\hat{\psi}_{\sigma}(\mathbf{r})$ is a field operator with spin projection $\sigma = \uparrow,\downarrow$ and $ v_0 < 0 $ is the bare interaction strength.
The contact interaction introduces ultraviolet divergences that are regularized in terms of the scattering length $a_\mathrm{2D}$ of the associated two-fermion problem in vacuum
\begin{equation}
\frac{1}{v_{0}} =- \int \frac{d\mathbf{k}}{(2\pi)^{2}} \, \frac{1}{{\frac{k^2}{m} }+\be},
\label{equ:2bvac}
\end{equation}
where $\be=1/ma_\mathrm{2D}^2$ is the binding energy of the two-body bound state (of radius $a_\mathrm{2D}$), which in a 2D vacuum exists for any strength of the bare interaction $v_0$ (see Supplemental Material of Ref.\til\cite{Bauer-2014} for a discussion of the regularization procedure).

In the present paper, we consider the strict zero-temperature limit, that is, we take the limit $T \to 0$  beforehand in all relevant equations obtained within the finite-temperature (Matsubara) formalism. Accordingly, all discrete fermionic $i\omega_n=i (2n+1)\pi T$ and bosonic $i\Omega_\nu=i 2\nu\pi T$ Matsubara frequencies are replaced by continuous frequencies $i\omega$ and $i\Omega$, and the corresponding summations, $T\sum_n$ and $T\sum_\nu$, are replaced by their continuous counterparts $\int d \omega/(2\pi)$ and $\int d \Omega/(2\pi)$, respectively.

For a system in which the two spin populations are imbalanced,  we find it convenient to introduce a polarization parameter 
$p=(n_\uparrow-n_\downarrow)/(n_\uparrow+n_\downarrow)$, which uniquely defines the  densities of the two species once the total density is fixed to the reference value $n=n_\uparrow+n_\downarrow=(\kF)^2/2 \pi$ of a non-interacting Fermi gas with Fermi momentum $\kF$ and Fermi energy $\eF = (\kF)^2/2 m$. Consequently, we define Fermi momenta $\kFs$ for each species through $n_\sigma=(\kFs)^2/4\pi$.

We also define an effective coupling parameter $ \gtwoD=\frac{1}{2}\ln\left(\be/2\eF\right)$ following previous studies on the BCS-BEC crossover in 2D   \cite{Bloom1975,Bertaina-2011}. The coupling parameter ranges from the weak-coupling regime $(\gtwoD \lesssim -2)$, where pairs are loosely bound $(\kF a_\mathrm{2D} \gg 1)$, to the strong-coupling regime $(\gtwoD \gtrsim 2)$ where pairs are tightly bound $(\kF a_\mathrm{2D} \ll 1)$, with $a_\mathrm{2D}$ measuring the radius of the pair.

While the main focus of the present paper is on two dimensions, for the sake of comparison we will also present some results obtained with the same theory in three dimensions. 
In this case, the coupling strength
is parametrized by $\gthreeD$ \cite{Strinati2018}, where $a_\mathrm{F}$ is the scattering length in 3D, and the opposite weak- and strong-coupling regimes correspond to $\gthreeD \lesssim -1$ and $\gthreeD \gtrsim 1$, respectively. 
\begin{figure}[t]
\centering
\includegraphics[width=0.95\linewidth]{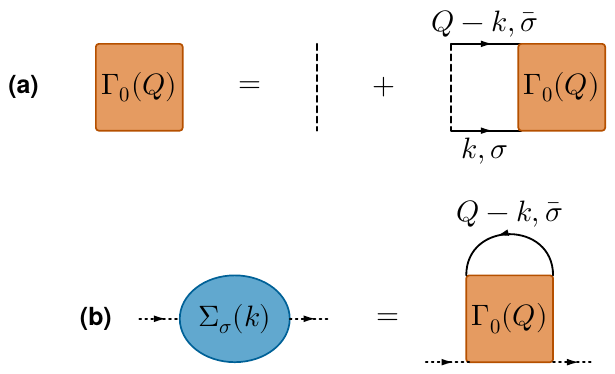}
\caption{Feynman diagrams for  (a) the particle-particle propagator $\Gamma_{0}(Q)$ and (b) self-energy $\Sigma_\sigma(k)$, with the notation $k=(\vk,i\omega)$, $Q=(\vQ,i\Omega)$, $\sigma=\downarrow,\uparrow$, $\bar{\sigma}=\uparrow,\downarrow$. Dashed lines: bare contact interaction $v_0$. Full lines with arrow: bare Green’s functions $G_{0\sigma}(k)$.}
\label{fig:feyndiag}
\end{figure}

\subsection{Non-self-consistent $t$-matrix approximation}
\label{sec:nsct}

The starting point of the diagrammatic many-body  approach adopted in the present work is the non-self-consistent $t$-matrix (NSCT) approximation, which has been widely used to describe the BCS–BEC crossover in ultracold Fermi gases and related systems~\cite{Chen2005,Strinati2018,Pini2019}. 

Under the hypothesis of a contact potential and translational invariance, the $t$-matrix or particle-particle (ladder) propagator $\Gamma_0(\vQ,i\Omega)$ satisfies the Bethe-Salpeter equation depicted in Fig.\til\ref{fig:feyndiag}(a), which admits the simple expression
\begin{equation}
\begin{aligned}
&\Gamma_0(\vQ,i\Omega) 
= v_0 -v_0 \Gamma_0(\vQ,i\Omega) 
\int \frac{d\mathbf{k^\prime}}{(2\pi)^2}  \\
&
\times \int_{-\infty}^{\infty}\frac{d\omega}{2\pi} \,
G_{0\uparrow}(\vQ-\mathbf{k^\prime}, i\Omega - i\omega) \,
G_{0\downarrow}(\mathbf{k^\prime}, i\omega),
\label{equ:gammaseries}
\end{aligned}
\end{equation}
where $G_{0\sigma}(\vk,i\omega)^{-1}=i\omega- \xi_\vk^\sigma$ with $\xi_\vk^\sigma = \frac{\vk^2}{2m}-\mu_\sigma$ is the bare Green's function and $\mu_\sigma$ the chemical potential of species $\sigma$.
After performing the integration over the internal frequency $\omega$,  Eq.\til\eqref{equ:gammaseries} can be solved straightforwardly yielding
\begin{equation}
\Gamma_0(\vQ, i\Omega)^{-1} = \frac{1}{v_0} + \int \frac{d\mathbf{k^\prime}}{(2\pi)^2} \, \frac{1 - \Theta(-\xi_{\vQ-\mathbf{k^\prime}}^{\uparrow}) - \Theta(-\xi_{\mathbf{k^\prime}}^{\downarrow})}{\xi_{\vQ-\mathbf{k^\prime}}^{\uparrow} + \xi_{\mathbf{k^\prime}}^{\downarrow} - i\Omega},
\label{equ:gammamats}
\end{equation}
where the contact interaction parameter $v_0$ is regularized through Eq. \eqref{equ:2bvac}.

The fermionic self-energy corresponding to the diagrammatic representation in Fig.\til\ref{fig:feyndiag}(b) is given by
\begin{multline}
\label{equ:se}
\Sigma_{\sigma}\left(\vk, i\omega\right)
    = \int \!\frac{d \vQ}{(2 \pi)^2} 
      \int \!\frac{d \Omega}{(2 \pi)} 
      \Gamma_0\!\left(\vQ,  i\Omega\right) \times \\ 
      G_{0\bar{\sigma}}\!\left(\vQ-\vk, 
      i\Omega-i\omega\right),
\end{multline}
where $\bar{\sigma}$ indicates the opposite of $\sigma$.
We note that, owing to rotational invariance, $\Gamma_0(\vQ, i\Omega)=\Gamma_0(|\vQ|, i\Omega)$ and $\Sigma_{\sigma}(\vk, i\omega)=\Sigma_{\sigma}(|\vk|, i\omega)$.  

The dressed single-particle Green's function $G_\sigma(\vk,\omega)$ is then obtained in terms of the above quantities through the Dyson equation
\begin{equation}
    G_\sigma(\vk,i\omega)=\left[ i\omega-\frac{k^2}{2m}+\mu_\sigma-\Sigma_\sigma(\vk,i\omega)\right]^{-1},
    \label{equ:gdress}
\end{equation}
and determines the total density of each component $n_\sigma$ as follows
\begin{align}
    n_\sigma&=\int \!\!\frac{d \vk}{(2 \pi)^2} 
      \int \!\frac{d \omega}{2 \pi}\, e^{i\omega0^+}\, 
      G_{\sigma}\!\left(\mathbf{k}, 
      i \omega\right) \label{equ:dens}\\
      &=\int \!\frac{d \vk}{(2 \pi)^2} \, n_{\sigma}(k).
\end{align}

For a given choice of the polarization $p$, the densities of both species are fully specified ($n_\uparrow/n=(1+p)/2$ and $n_\downarrow/n=(1-p)/2$), and the respective chemical potentials $\mu_\uparrow$ and $\mu_\downarrow$ are obtained by solving the two  equations\til\eqref{equ:dens} once the coupling strength $\gtwoD$ is fixed to a given value.

\subsection{Minimal self-consistent $t$-matrix approximation}
\label{sec:MSCT}

The Luttinger theorem \cite{Luttinger1960} asserts that, at zero temperature, the volume enclosed by the FS is determined solely by the particle density and remains unaffected by interactions. In the case of a polarized Fermi gas, this implies that the FS of each spin component must enclose a volume equal to that of a non-interacting gas with the same density. 

While the theorem was originally proved for the exact many-body theory, its validity within approximate schemes is not guaranteed. Notably, it has been shown\til\cite{Pieri2017} that conserving and fully self-consistent diagrammatic approximations—such as the Luttinger-Ward (or self-consistent 
$t$-matrix) approach—do preserve the Luttinger theorem for both balanced and imbalanced systems, while approximate schemes may even strongly  violate it\til\cite{Urban2014,Pantel2014,Pini2023}. 
One such scheme is the NSCT approximation,
which is known to produce unphysical momentum distributions and spin susceptibilities when a Fermi gas is polarized \cite{Urban2014,Pantel2014,Pini2023}. 

In order to remedy this shortcoming and  work towards self-consistency, we extend the NSCT scheme by including a minimal degree of self-consistency as follows. We recall that the interacting Fermi momentum $|\vk|=\kL $ is defined by $G(\vk,0)^{-1}=0$\til\cite{Luttinger1960}, that is
\begin{equation}
\label{equ:lutteq}
\mu_\sigma = \frac{\kLsq}{2m}
+\Sigma_\sigma\!\left(\mathbf{k} = \vkL,\, i\omega = 0\right),
\end{equation}
and identifies the position  of the discontinuity in the momentum distribution of a Fermi liquid (FL).
In an isotropic system the Luttinger theorem is satisfied if $\kL=\kFs$. 

If we consider dressing the bare propagator $G_{0\sigma} \left(\vk,i\omega\right)$  with the constant self-energy $\Sigma_\sigma( \kL, 0 )$ appearing in Eq.\til\eqref{equ:lutteq}, said propagator is seen to maintain a non-interacting-like form 
\begin{align}
\label{equ:g0g0til}
    G_{0\sigma} (\vk,i\omega)^{-1} \rightarrow \tilde{G}_{0\sigma}(\vk,i\omega)^{-1}&=i\omega-\tilde\xi_\vk^\sigma, \\
    &=i\omega-\frac{\vk^2}{2m}+\muL, 
\end{align}
where we define an effective  chemical potential $\muL\equiv \mu_\sigma- \Sigma_\sigma( \kL, 0 )$.
As a result, the interacting Fermi momentum $\kL$ is simply given by $\kL=\sqrt{2m\muL}$ according to Eq. \eqref{equ:lutteq}.

A new self-energy $\tilde{\Sigma}_{\sigma}(\vk,i\omega)$ can then be defined 
through Eqs.\til\eqref{equ:gammaseries} and\til\eqref{equ:se} after making the replacement\til\eqref{equ:g0g0til} 
therein. To maintain self-consistency,  the constant shift $\Sigma_\sigma( \kL, 0 )$
must be replaced with $\tilde\Sigma_\sigma( \kL, 0 )$, that is $\mu_\sigma \rightarrow \muL = \mu_\sigma- \tilde\Sigma^0_\sigma$ with
\begin{equation}
\tilde\Sigma^0_\sigma\equiv\tilde\Sigma_\sigma( \kL, 0 ).    
\label{equ:s0til}
\end{equation}

The Dyson equation defining the dressed Green's function \eqref{equ:gdress} may be recast as follows,
\begin{align}
G_{\sigma}(\vk,i\omega) &= \left[{G}_{0\sigma}^{-1}(\vk,i\omega) - \tilde \Sigma_{\sigma}(\vk,i\omega)\right]^{-1} \\
&= \left[\tilde{G}_{0\sigma}^{-1}(\vk,i\omega) - \left(\tilde \Sigma_{\sigma}(\vk,i\omega)-\tilde\Sigma^0_{\sigma}\right)\right]^{-1},
\label{equ:dysMSCT}
\end{align}
where $\tilde\Sigma^{\rm eff}_\sigma(\vk,i\omega)\equiv \tilde \Sigma_{\sigma}(\vk,i\omega)-\tilde\Sigma^0_{\sigma}$ can be interpreted as an effective self-energy renormalizing the bare-like (mean-field shifted) Green's function $\tilde{G}_{0\sigma}(\vk,i\omega)$.
The above equations define a minimal self-consistent $t$-matrix (MSCT) scheme,   whereby the chemical potentials $\muL$ continue to be determined through Eq.\til\eqref{equ:dens} by means of  Eq. \eqref{equ:dysMSCT}.

The MSCT FS is defined by $G_\sigma(\vk,0)^{-1}=0$ thus implying, according to Eq. \eqref{equ:dysMSCT}, that  $\tilde{G}_{0\sigma}(\vk,0)^{-1}=0$. Therefore, the underlying minimally dressed system, identified by $\tilde{G}_{0\sigma}(\vk,i\omega)$, and the fully dressed system, identified by $G_{\sigma}(\vk,i\omega)$, share the same Fermi momentum $\vkL$. 
This non-trivial property of the MSCT scheme turns out to be the crucial ingredient able to remove  the aforementioned unphysical effects arising in the NSCT scheme. 

It should be noted that the same self-consistent shift $\tilde{\Sigma}_0$ within a NSCT approximation was first introduced in Refs.~\cite{Perali2002,Pieri2004} for a balanced (unpolarized) Fermi gas. For the polarized Fermi gas, a similar shift was used in Ref.~\cite{Pantel2014} (with the minor difference that $\tilde{\Sigma}_0$ was calculated at $k_{{\rm F}\sigma}$ rather than  at $\tilde{k}_{{\rm F\sigma}}$).

\subsection{Generalized Thouless criterion and critical polarization}

At zero temperature and zero polarization the gas is in the condensed state\til\cite{Randeria-1989,Bertaina-2011,Levinsen-2015,Salasnich-2015,2015-He,Boettcher-2016,Vitali-2017,Zielinski-2020,Sobirey-2021,VanLoon-2023}.
For large enough spin imbalance (with the convention $n_\uparrow > n_\downarrow$)  a transition to the normal state takes place such that the static pairing susceptibility of  the normal state becomes divergent. At the simplest level of approximation, the latter quantity   is identified  by $-\Gamma_0( \vQ, i\Omega=0)$\til\cite{Pini2021}, and offers  a generalized form of the Thouless criterion
\begin{equation}
\label{equ:Thoulcr}
\left[\left.\Gamma_0( \vQ, i\Omega= 0)\right|_{p=p_c}\right]^{-1} = 0.   
\end{equation}
When Eq.~\eqref{equ:Thoulcr} is satisfied at a finite pair momentum $|Q|=\sQfflo\neq 0$, it indicates a transition to the FFLO superfluid phase \cite{Fulde1964,Larkin1964}, while a solution at $Q=0$ corresponds to a transition to a polarized BCS superfluid \cite{Sarma1963}. 
For a given choice of the chemical potentials ($\mu_\sigma$ for NSCT, $\tilde\mu_\sigma$ for MSCT) appearing in the fermionic propagators, condition\til\eqref{equ:Thoulcr} determines simultaneously the critical polarization $p_c$ and the momentum of the Cooper pairs \cite{Strinati2018}.

For a given choice of coupling $\gtwoD$, Eqs.~\eqref{equ:dens} and  \eqref{equ:Thoulcr} represent a set of equations to be solved for the unknowns $p_c$ and $\mu_\uparrow$, $\mu_\downarrow$ (in the NSCT approximation) or $\tilde\mu_\uparrow$,$\tilde\mu_\downarrow$ (in the MSCT approximation).  In this way, the critical polarization $p_c$ separating the normal and superfluid phases can be obtained as a function of the coupling strength.

\section{Numerical Implementation}
\label{sec:NumImpl}

To compute the self-energy\til\eqref{equ:se} we introduce the spectral representation of the pair propagator $\Gamma_0\left(\vQ, i\Omega\right)$
\begin{equation}
\Gamma_0\!\left(\vQ, i\Omega\right)
   = -\int \!\frac{d\Omb}{\pi}\,
     \frac{\IM \Gamma_0^\mathrm{R}(\vQ, \Omb)}
          {i \Omega-\Omb},
\label{equ:GammaSpectral}
\end{equation}
where $\Gamma_0^\mathrm{R}(\vQ, \Omb) \equiv 
\Gamma_0(\vQ, i \Omega \!\to\! \Omb+i 0^+)$ 
is the retarded pair propagator. In appendix \ref{sec:ImGammaRet} we provide a comprehensive illustration of the structure (poles and two-particle continuum) of the pair spectral weight function ${\rm Im} \Gamma_0^{\rm R}(\mathbf Q,\bar{\Omega})$. 
Substituting it into Eq.~\eqref{equ:se} and computing the integral over  $\Omega$, the self–energy becomes
\begin{multline}
\label{equ:sigspectr}
\Sigma_{\sigma}\!\left(\mathbf{k}, i \omega\right)
   = - \!\int \!\frac{d \vQ}{(2 \pi)^2}
      \int \!\frac{d\Omb}{\pi} \,
      \IM \Gamma_{0}^{\rm R}(\vQ, \Omb)  \\
   \quad \times 
      \frac{f( \xi_{\vQ-\vk}^{ \bar{\sigma}})+b(\Omb) }
           {i \omega-\Omb+ \xi_{\vQ-\vk}^{\bar{\sigma}}  },
\end{multline}
where $f(x)=\Theta(-x)$ and $b(x)=-\Theta(-x)$ are the  Fermi  and Bose functions in the zero-temperature limit, respectively.

Eq.\til\eqref{equ:sigspectr} can be split into two terms
$\Sigma_{\sigma}(\vk, i\omega)
   = \Sigma_{\sigma}^{f}(\mathbf{k}, i \omega) 
   + \Sigma_{\sigma}^{b}(\mathbf{k}, i \omega)$, 
defined by
\begin{align}
    \Sigma^f_{\sigma}\!\left(\vk,i \omega\right)
   &= - \int \frac{d \vQ}{(2 \pi)^2}
      \int \frac{d\Omb}{\pi} \,
        \frac{ \IM \Gamma_{0}^{\rm R}(\vQ, \Omb) }
           {i \omega-\Omb+ \xi_{\vQ-\vk}^{\bar{\sigma}}  } \, f( \xi_{\vQ-\vk}^{ \bar{\sigma}}) , \\
    \Sigma^b_{\sigma}\!\left(\vk,i \omega\right)
   &= - \int \frac{d \vQ}{(2 \pi)^2}
      \int \frac{d\Omb}{\pi} \,
        \frac{\IM \Gamma_{0}^{\rm R}(\vQ, \Omb) }
           {i \omega-\Omb+ \xi_{\vQ-\vk}^{\bar{\sigma}}  } \, b(\Omb).
\label{equ:sigfb}
\end{align}
In the $f$ term the integral over $\Omb$  can be performed analytically taking advantage of the spectral representation\til\eqref{equ:GammaSpectral}
in the reverse sense with $i\Omega\to i\omega+\xi_{\vQ-\vk}^{\bar{\sigma}} $, thus yielding
\begin{equation}
\begin{split}
\Sigma_{\sigma}^{f}(\mathbf{k}, i \omega) 
   =\int \!\frac{d \vQ}{(2 \pi)^2} \,
     f(\xi_{\vQ-\vk}^{\bar{\sigma}})  \, \Gamma_0\!\left(\vQ, 
     i \omega+\xi_{\vQ-\vk}^{\bar{\sigma}}\right).
\end{split}
\end{equation}

The $b$ term is  subdivided into two further terms, based on the structure of the pair spectral weight function  $\IM \Gamma_{0}^{\rm R}(\vQ, \Omb) $ (see appendix \ref{sec:ImGammaRet}). 
The first subterm stems from the  poles of the pair propagator, which determine the collective modes of the system and are associated with $\delta$-like 
contributions to $\IM    \Gamma_0^{\rm R}$: 
\begin{align}
\IM \Gamma_0^{\rm R}(\vQ, \Omb)  
   &=  -\pi W_l(Q) \delta(\Omb-\Omb^{*}_l(Q)),
\end{align}
where $\Omb^*_l(Q)$ is the pole dispersion,
$W_l(Q) = \left[\partial_\Omb \RE \Gamma^{-1} (\vQ,\Omb)\right]^{-1}_{\Omb=\Omb^*_i(Q)}$
its spectral weight, and $l$ ranges over the number of dispersions that are present for a given momentum $Q$. The number of these dispersion can vary from 0 to 2, depending on $Q$. In particular, at negative frequencies, only one dispersion appears, and in a small range around $Q=0$.   Note also that we have used rotational invariance in the expression above, with
$ Q\equiv|\vQ|$.
The second subterm  originates instead from the two-particle continuum occurring for $\Omb > \Omb_{\rm th}(Q)$.

Therefore, the $b$ term now reads
\begin{equation}
\begin{split}
&\Sigma_{\sigma}^{b}(\mathbf{k}, i \omega)
   = \!\int \!\frac{d \vQ}{(2 \pi)^2} \, \sum_l
       \frac{ W_l(Q) b(\Omb^*_l(Q)) }
            { i\omega-\Omb^*_l(Q)+\xi_{\vQ-\mathbf{k}, \bar{\sigma}} } \\[6pt]
     &- \!\int \!\frac{d \vQ}{(2 \pi)^2} 
        \int_{\Omb_{\rm th}(Q)}^{+\infty} \!\frac{d\Omb}{\pi}\,
        \IM \Gamma_0^{\rm R}(\vQ, \Omb)\,
        \frac{b(\Omb)}
             {i \omega - \Omb + \xi_{\vQ-\vk, \bar{\sigma}}} ,
\end{split}
\label{numerical}
\end{equation} 
with $\Omb_{\rm th}(Q)$ the threshold of the two-particle continuum. In addition, information on further internal thresholds occurring above $\Omb_{\rm th}(Q)$ (see Sec.~\ref{subsec-pair} and appendix \ref{sec:ImGammaRet}) is used when performing the numerical integrals in the second term of Eq.~\eqref{numerical}. We also notice that, because of the Bose function, only the dispersion at negative frequencies matters in the first term in Eq.~\eqref{numerical} (and actually with a negligible contribution because of its small weight and momentum range).

Once the self-energy is computed, the momentum distribution function $n_\sigma(\vk)$ is obtained by evaluating the integral
over the frequency $\omega$ in Eq.\til\eqref{equ:dens}. To do so we introduce a large energy cutoff $\omega_c^{(1)} \simeq 100 \eF$ beyond which the following asymptotic behaviors hold\til\cite{Pisani2025}
\begin{align}
    \Sigma^{f}_{\sigma}(\mathbf{k}, i \omega)&\xrightarrow{\omega \gtrsim \omega_c^{(1)}} n^0_{\mu_\mathrm{F}} \Gamma_0(\mathbf{k}, i \omega), \\
     \Sigma^{b}_{\sigma}(\mathbf{k}, i \omega) &\xrightarrow{\omega \gtrsim \omega_c^{(1)}} -   \Delta_{\infty}^2 G_0(-\mathbf{k}, -i\omega),
\label{eq:asymptoticSigma}
\end{align}
where $n^0_{\mu_\mathrm{F}} = \int \frac{d^2k}{(2 \pi)^2}f(\xi_{\vk})$ and
\begin{equation}
\Delta^2_{\infty} = -\int \frac{d \vQ}{(2 \pi)^2} \int \frac{\mathrm{d} \Omb}{\pi} \operatorname{Im} \Gamma_0^{\mathrm{R}}(\vQ, \Omb) b(\Omb)    
\end{equation}
defines the Tan's Contact through $C=\Delta^2_\infty  / m^2$ \cite{Strinati2018}. 

To avoid numerical instabilities deriving from the pole of the dressed Green's function (which locates the Fermi step of $n_\sigma(\vk)$), we add and subtract the associated quasi-particle Green's function so that the momentum distribution in Eq. \eqref{equ:dens} is recast as
\begin{align}
    n_\sigma(\vk)&=\int_{-\infty}^{\infty} \frac{d \omega}{2 \pi}\, e^{i\omega0^+}\,  G_{\sigma}\!\left(\mathbf{k}, 
      i \omega\right) \nonumber\\
      &=2 \int_{0}^{\omega_c^{(2)}}\!\!\frac{d \omega}{2 \pi}\,  
      \RE\left[G_{\sigma}(\mathbf{k}, i \omega)- G^\mathrm{qp}_{\sigma}(\mathbf{k}, i \omega) \right]  \nonumber \\
      & + (1-Z_\sigma)/2 + Z_\sigma \,  \Theta(\kL-k),
      \label{eq:densitynum}
\end{align}
where
\begin{align}
    G^\mathrm{qp}_{\sigma}(\mathbf{k}, i \omega) &=
     \left[ \frac{i\omega}{Z_\sigma}- \frac{k^2}{2m}+\frac{\kLsq}{2m} \right]^{-1}, \\
     Z_\sigma^{-1}&=\left. 1-\frac{\partial \IM \Sigma_\sigma(\vkL,i\omega)}{\partial \omega } \right|_{\omega=0^+} ,
     \label{equ:Zmats}
\end{align}
with $\vkL$ determined by the equation $G_\sigma(\vkL,0)^{-1}=0$. 

The cutoff $\omega_c^{(2)}$ in Eq.~\eqref{eq:densitynum} is taken sufficiently large to approximate $G_{\sigma}(\mathbf{k}, i \omega) \simeq 1/i \omega$ and $G^\mathrm{qp}_{\sigma}(\mathbf{k}, i \omega) \simeq Z_\sigma/i \omega$ for $|\omega| > \omega_c^{(2)}$. Therefore, the integral of the difference between $G$ and $G^\mathrm{qp}$, multiplied by the convergence factor $e^{i \omega 0^+}$ can be performed analytically in this large frequency region and yields the term  $(1-Z_\sigma)/2$ appearing in Eq.~\eqref{eq:densitynum}.

\section{critical polarization} 
\label{sec:Results}

In this section we present numerical results for the critical polarization $p_c$ as a function of coupling strength $\gtwoD$, obtained by solving simultaneously Eqs.\til\eqref{equ:dens} and Eq.\til\eqref{equ:Thoulcr}. 
In practice the solution of this set of three equations can be significantly simplified by taking advantage of the fact that the inverse pair susceptibility appearing in the Thouless condition \eqref{equ:Thoulcr} acquires the analytical form (see Eq.\eqref{equ:gammaThoul})
\begin{multline}
-\Gamma_0(|\vQ|=\sQfflo,i\Omega= 0)^{-1} \\
=\frac{m}{4\pi}\times
      \begin{cases}    
      \ln \frac{ \left(\sqrt{2m\mu_\uparrow}-\sqrt{2m\mu_\downarrow}\right)^2 }{4m\be}  &\text{if} \;\sQfflo \neq 0  \\ 
      \ln{ \frac{|\mu_\uparrow-\mu_\downarrow|}{\be} } &\text{otherwise},
\end{cases}
\label{equ:Thoulequ}
\end{multline}
for the NSCT approximation, while for the MSCT approximation the chemical potentials $\mu_\uparrow$,$\mu_\downarrow$
are replaced by the shifted chemical potentials $\tilde\mu_\uparrow$, $\tilde\mu_\downarrow$.

In essence,  this set of three coupled  equations is reduced to solving the condition $n_\uparrow+n_\downarrow=n$ (with $n_\sigma$ given by Eq.\til\eqref{equ:dens}) in one of the two unknown chemical potentials,  
the other being uniquely determined by Eq.\til\eqref{equ:Thoulequ}. The critical polarization $p_c$ is then extracted by inserting the computed densities $n_\sigma$ into its definition $p_c=(n_\uparrow-n_\downarrow)/(n_\uparrow+n_\downarrow)$.

\subsection{Non self-consistent $t$-matrix approximation} 
\label{sec:nsctnum}
\begin{figure}
    \centering
\includegraphics[width=0.95\linewidth]{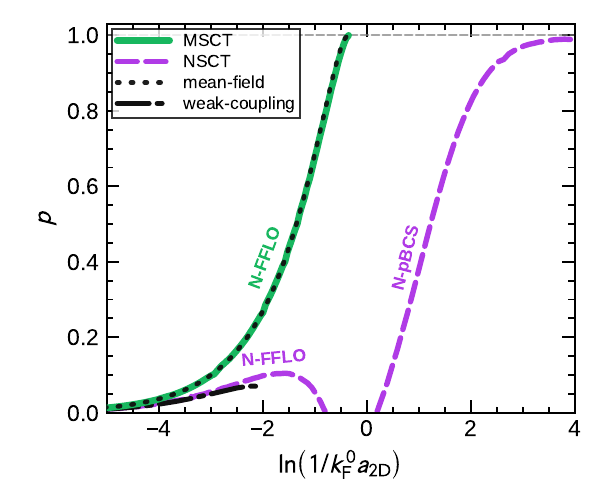}
    \caption{2D critical polarization $p=p_c$ versus coupling $\gtwoD$ within the NSCT (dashed line) and MSCT (solid line)  approaches. Dotted line:  mean-field critical polarization. Dash-dotted line: weak-coupling expression\til\eqref{equ:wclim}. For $p > p_c$ the system is in the normal phase. The labels (N-FFLO or N-pBCS) indicate the nature of the transition at the critical polarization.}
    \label{fig:phasediag}
\end{figure}

The resulting critical polarization is reported in Fig.\til\ref{fig:phasediag}. In particular, the critical polarization $p_c $ versus coupling $\gtwoD$  within the NSCT approximation is reported  as a dashed line and is seen to be composed of two branches. 
The branch with $\gtwoD<0$ represents a transition to an FFLO state, with the pair susceptibility diverging at a finite pairing momentum $\sQfflo \equiv \sqrt{4m\be}$
(see Eqs. \ref{equ:qchc}).   
In the weak coupling limit $(\gtwoD \lesssim -2 )$ the critical polarization approaches the perturbative expression, 
\begin{equation}
p_c \simeq \left( 2 + 3/g\right) e^g
\label{equ:wclim}
\end{equation}
obtained at leading order in $1/g$ in appendix \ref{sec:wclim}, where $g\equiv\gtwoD$.

The second branch has $\gtwoD>0$ and the transition is to a polarized BCS superfluid (pBCS), also called the Sarma  state\til\cite{Strinati2018}, whereby bosonic pairs condense with zero center-of-mass momentum in the presence of a polarized gas of residual unpaired fermions.
At coupling values between these two branches (around $\gtwoD \simeq 0$), the transition to the condensed state unphysically disappears,
and the gas remains unexpectedly in the normal phase
 for all values of the polarization.
 
\begin{figure}[t]
    \centering
    \includegraphics[width=0.95\linewidth]{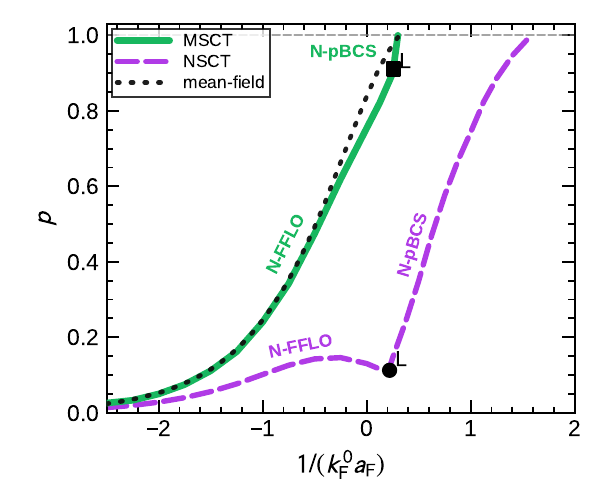}
    \caption{3D critical polarization $p=p_c$ versus coupling $(k_\mathrm{F}a_\mathrm{F})^{-1}$ within the NSCT (dashed line) and MSCT (solid line) approaches. Dotted line: mean-field critical polarization. The position of the Lifhsitz points L within the two approaches is also indicated. For $p > p_c$ the system is in the normal phase. The labels (N-FFLO or N-pBCS) indicate the nature of the transition at the critical polarization.}
    \label{fig:phasediag3D}
\end{figure}

To gain insight into the above scenario,
we have computed the NSCT critical line for the corresponding  3D homogeneous system, following the same NSCT scheme (see also \cite{Tartari2011thesis}).
The outcome is shown in Fig.\til\ref{fig:phasediag3D} as a dashed line made of two branches that meet  at the Lifshitz point L ($\gthreeD=0.22$), where the transition changes nature from FFLO to polarized BCS superfluid \til\cite{Pini2023}. 
In the 2D system the L point is absent and effectively replaced by a coupling range where the transition ceases to exist ($-0.8 \lesssim \gtwoD \lesssim 0.2$). 

\begin{figure}
    \centering
     \includegraphics[width=0.95\linewidth]{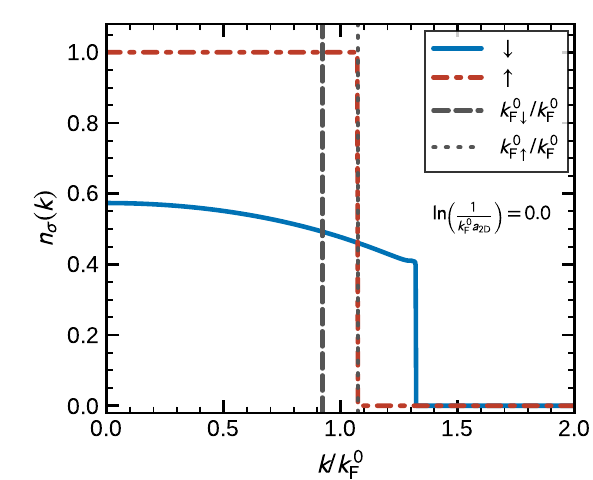}
    \caption{Momentum distribution function $n_\sigma(k)$ within the NSCT approximation for 
    $\gtwoD =0$ and polarization $p=0.15$. The
   vertical  lines represent the non-interacting Fermi momenta $k^0_{\mathrm{F}\downarrow}$ (dashed line) and $k^0_{\mathrm{F}\uparrow}$ (dotted line). The
   interacting Fermi momenta $\kLdn$ and $\kLup$ can be read by the positions of the discontinuities in the respective distributions.}
    \label{fig:nkNSC}
\end{figure}

In Fig.\til\ref{fig:phasediag} the NSCT approximation (dashed line) predicts, in the  region approaching the coupling $\gtwoD = -0.8$, a non-physical decrease of the critical polarization, implying that the ordered phase becomes less and less favorable as the coupling increases. A similar but less dramatic effect is also found in the 3D system, just before the Lifshitz point L (dashed line in Fig.\til\ref{fig:phasediag3D} for $-0.3 \lesssim \gthreeD \le 0.22$).
To clarify the  origin of this anomalous behavior, we examine the momentum distribution functions of both species $n_\sigma(k)$ of the 2D system, as reported in Fig.\til\ref{fig:nkNSC} for the coupling $\gtwoD=0$ and polarization $p=0.15$. We observe that the majority species distribution appears to be that of a non-interacting system with $n_\uparrow(\vk)\equiv\Theta(k_{\mathrm{F}\uparrow}-k)$. This is because the minority chemical potential $\mu_\downarrow$ is renormalized to negative values and, 
when inserted in the bare Green's function $G_{0\downarrow}(\vQ-\vk,i\Omega-i\omega)$, yields a vanishing value of the convolution appearing in 
Eq.~\eqref{equ:sigfb} for the self-energy with $\sigma=\uparrow$. 

In practice, within the NSCT scheme, the  propagators $G_{0\sigma}(\vk,i\omega)^{-1}=i\omega-\xi^\sigma_\vk$ (solid lines in Fig.\til\ref{fig:feyndiag}) appearing in the self-energy convolution 
\eqref{equ:se} are {\em bare} ones; as such, they do not take into account any consistent renormalization of the related Fermi momenta, resulting in unphysical outcomes as shown above for the majority species.
This lack of self-consistency was recognized in previous work as the root of the problems with the NSCT approximation when applied to polarized Fermi gases \cite{Schneider2009,Pantel2014,Urban2014, Pini2023}, and also when calculating the spin-spin correlation function of the balanced system\til\cite{Pantel2014,Takashi2012}.
Notably, in Ref.\til\cite{Pini2023} it was pointed out that the NSCT approach strictly violates the Luttinger theorem\til\cite{Luttinger1960, Luttinger2} in a 3D system, whereas full self-consistency was shown to restore the validity of the theorem.

An analogous violation is found here for the 2D system, as illustrated in Fig.\til\ref{fig:nkNSC}.
On one hand, when $\mu_{\downarrow} < 0$ the majority species is found to be  non-interacting, with the Luttinger theorem trivially respected $\kLup=k^0_{\mathrm{F}\uparrow}$ (dotted line). On the other hand, the discrepancy between the position $\kLdn$ of the Fermi momentum in $n_\downarrow(k)$ and the Fermi momentum $k^0_{\mathrm{F}\downarrow}$ of the non-interacting system (dashed line) clearly illustrates the violation of the Luttinger theorem for the minority species.

On the basis of the above analysis, we expect that, whereas full self-consistency is able to remedy the aforementioned pathologies both qualitatively and quantitatively, the adoption of a minimal degree of self-consistency should be sufficient to restore at least qualitatively the correct physical picture by introducing a consistent treatment of the Fermi momenta.  
This is what the MSCT scheme introduced in sec.~\ref{sec:MSCT} is expected to do, as  will be shown in the next section. 

Before concluding we point out another important difference between the 2D and the 3D systems, that is the absence or presence, respectively, of the polaron-molecule transition at $p_c=1$.
In the 3D case, Fig.\til\ref{fig:phasediag3D}  shows that the critical value $\gthreeD=1.6$ for the polaron-molecule transition expected within the NSCT approximation \cite{Combescot2009,Punk2009,Fratini2012} is recovered in the $p\to 1$ limit.
On the contrary, in Fig.\til\ref{fig:phasediag} we find that the critical polarization approaches the polaronic limit $p_c=1$ only asymptotically, as also found in a recent NSCT study of 2D Bose-Fermi mixtures\til\cite{Pisani2025} in the limit of a vanishing concentration of bosons. As argued in that work, the NSCT approach fails to capture the polaron-molecule transition in 2D, as it does not take into account three-body correlations, that were originally demonstrated in \cite{Parish2011} via a variational approach  to be essential for a physically sound description of the molecular phase in 2D.
 
\subsection{Minimal self-consistent $t$-matrix approximation}
\label{sec:NSC+MSF}

\begin{figure}[t!]
\centering
\includegraphics[width=0.95\linewidth]{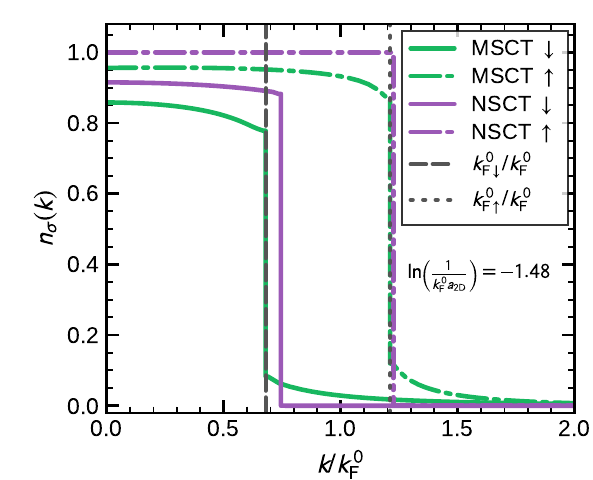}
    \caption{Momentum distribution function $n_\sigma(k)$ within the MSCT (green) and NSCT (magenta) approaches  for the majority (dot-dashed lines) and minority (solid lines) species at  $\gtwoD=-1.48$ (corresponding to $\varepsilon_0/E^0_{\rm F}=0.105)$ and $ p=0.5$.}
    \label{fig:MSCTvsnsct}
\end{figure}

To remedy the shortcomings of the NSCT approach just discussed,  we introduce a minimal degree of self-consistency as  outlined in sec.\til\ref{sec:MSCT} for the MSCT approach.
Figure~\ref{fig:MSCTvsnsct} presents a comparison between the two schemes for the momentum distributions. We focus on a case in which the NSCT approach clearly violates the Luttinger theorem for the minority species and unphysically predicts a non-interacting momentum distribution for the majority species. 
One sees that the MSCT scheme resolves both problems:  the momentum distributions now fulfill the Luttinger theorem almost exactly and the majority species  is now interacting.

The impact of this improvement of the theory on the critical polarization is shown in Fig.\til\ref{fig:phasediag}.
We see that a physically sensible behavior is now restored for the critical polarization (solid line): the tendency of the system to form condensed pairs increases with the coupling strength. In addition, in the single-impurity limit $(p\to 1$) the polaron-to-molecule transition is recovered. 
Finally, as in the NSCT approximation, the Lifshitz point is absent, but now this does not occur with an unphysical region separating two branches. Rather, the single transition line always remains towards an FFLO superfluid phase.

However, a striking feature of Fig.\til\ref{fig:phasediag} is that the MSCT critical curve is essentially indistinguishable from the corresponding mean-field (MF) critical curve (dotted line). 
This finding can be understood by noticing that if the MSCT approach exactly satisfied the Luttinger theorem, then its critical line would be bound to coincide with the MF critical line. 

This is because the MF critical curve is obtained by the Thouless criterion \eqref{equ:Thoulcr} with 
the chemical potentials $\mu_\sigma$ entering Eq.~\eqref{equ:Thoulequ} for $\Gamma_0(Q_{\rm FF},i\Omega=0)$ taken at their non-interacting values: $\mu_\sigma=\eFs$.
For the MSCT approach, instead, the shifted chemical potentials $\mut_\sigma$ enter the same equation. On the other hand, fulfillment of the Luttinger theorem requires $\mu_\sigma=\eFs+\tilde{\Sigma}(\kFs,0)$ and $\kL=\kFs$,  thus implying $\mut_\sigma=\mu_\sigma- \tilde\Sigma_\sigma( \kL, 0 )=\eFs$ and the full equivalence between the two equations. (We stress that this does not imply that the same equivalence holds for a generic theory that exactly obeys the Luttinger theorem, since the proof of the equivalence relies on the specific form of the Thouless criterion within the MSCT approximation.)

As a matter of fact,  we have verified that along the critical line, $\kLup$  differs from $k_{\rm F\uparrow}^0$ by 0.3\%, at most. Deviations from the Luttinger theorem are more significant for the minority species. 
However, they are confined to the region of large polarization where $\kLdn$ and $k_{\rm F\downarrow}^0$ both go to zero, so that the impact of these deviations on the critical polarization remains very small. Specifically, we have verified that the relative difference between $\kLdn$ and $k_{\rm F\downarrow}^0$ remains below 3\% for polarization $p < 0.7$ and only for polarization $p > 0.9$ does the relative difference exceeds 10\%.
As a consequence, the critical line is essentially unchanged with respect to the MF one.

A similar behavior is observed in the 3D case presented in Fig.\til\ref{fig:phasediag3D}, even though in this case some deviations between the MSCT and MF curves are evident for $p > 0.6 $. This is because in this case the relative difference between $\kLdn$ and $k_{\rm F\downarrow}^0$ exceeds 10\% already at $p \simeq 0. 65 $, that is, for polarizations at which $\kLdn$ and $k_{\rm F\downarrow}^0$ are still significant.
We remark that our results for the MSCT curve in 3D agree very well with the results obtained in \cite{Durel2020} within the approach called RPA$(\infty)$. Formally, the RPA$(\infty)$ approach sums the same ladder diagrams as in the NSCT approach, but uses the  $T=0$ diagrammatic theory formulated in the canonical ensemble \cite{Fetter-2003}, in which the bare propagators have by construction the discontinuity at $\kFs$. In this way, it basically coincides with our MSCT approach, which is instead formulated in the grand-canonical ensemble and requires the introduction of the self-energy shift $\tilde{\Sigma}_0$ to align the jump in the bare-like propagators $\tilde{G}_0$ with that of the interacting Green's function $G$. When $\kL =\kFs$, the two theories coincide. Note that the importance of introducing such a shift to match the grand-canonical perturbation theory with the canonical one was pointed out long time ago by Luttinger and Ward \cite{LuttingerWard-1960}.

As a side remark, we notice that in the 3D case the Lifshitz point in the MSCT approach is shifted to much higher polarizations compared with the NSCT approach, although its position remains almost unchanged at $\gthreeD=0.255$. The polaron-to-molecule transition at $p \to 1$ is pushed to lower values of the coupling strength.

\begin{figure}[t!]
    \includegraphics[width= 0.95\linewidth]{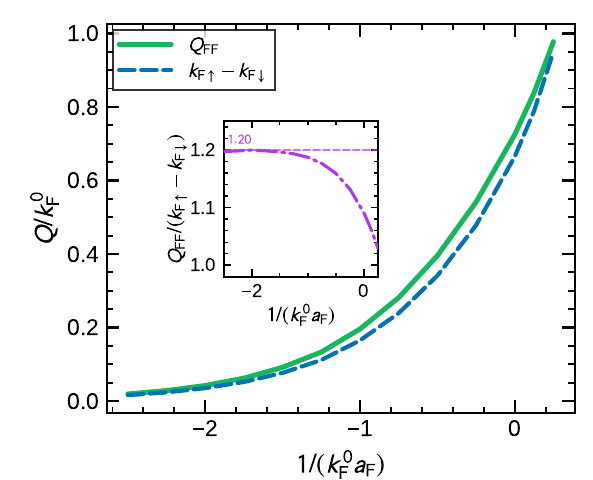}
    \caption{Comparison between the curve $Q=\sQfflo$ (full line) and the curve $Q=\kLup-\kLdn$ (dashed line) in 3D at $p=p_c$, as a function of the 3D coupling strength $\gthreeD$.
    Inset: the ratio $\sQfflo/(\kLup-\kLdn)$ (full line) is compared with its mean-field value $1.20$ in the weak-coupling limit \cite{Takada-1969}.} 
\label{fig:QffloVScoupl}
\end{figure}

Finally, within the MSCT approach,  it is interesting to quantify the relation between $\sQfflo$ and $\kLup-\kLdn$ in 3D  which, in contrast to its 2D counterpart, does not have an analytical form. This information will be useful for the discussion on criticality in the next section.
In Fig. \ref{fig:QffloVScoupl} we show the critical pair momentum $\sQfflo$ and the quantity $\kLup-\kLdn$ versus coupling in 3D: these two quantities are found to be identical in 2D but not in 3D.  This implies that the vectors $\vkLup$, $\vkLdn$, and $\vQfflo$ are collinear in 2D but not in 3D, as we illustrate in the next sections.
In the inset of Fig. \ref{fig:QffloVScoupl} we show that the ratio $\sQfflo/(\kLup-\kLdn)$ reaches its Fulde-Ferrell mean-field value 1.2  in the weak-coupling limit \cite{Takada-1969}. 

\section{Non-Fermi-liquid behavior}

In this section we examine the signatures of the quantum critical behavior through key quantities like the quasi-particle decay rate,  the quasi-particle weight $Z_\sigma$ at the Fermi step, the momentum distribution function $n_\sigma(k)$ and the bosonic and fermionic dynamical exponents. We focus on the MSCT scheme, which overcomes the  shortcomings of the NSCT approach pointed out above. (To avoid overburdening the notation we omit the tilde symbol over $\Sigma$ and $\Gamma_0$ indicating  the MSCT approximation.)

It is convenient to consider the imaginary part of the retarded self-energy, which after analytic continuation $i\omega \rightarrow \omb+i0^+$ of expression \eqref{equ:sigspectr} acquires the following form 
\begin{multline}
\IM\Sigma_{\sigma}^\mathrm{R}(\vk, \omb) = \int \frac{d{\bf k}^{\prime}}{(2\pi)^2} \IM \Gamma_0^\mathrm{R}(\vk^{\prime}+\vk, \omb + \xit_{\vk^{\prime}}^{\bar\sigma})\\
\times \left[ \Theta(-\xit_{\vk^{\prime}}^{\bar\sigma})-\Theta(-\omb -\xit_{\vk^{\prime}}^{\bar\sigma})\right].
\label{equ:imsigret0}
\end{multline}

We recall that in a Fermi liquid (FL) the lifetime $\tau_{\vk, \sigma}$ of a quasi-particle of given momentum $\vk$  and spin $\sigma$ is given by \cite{Varma2002}
\begin{equation}
    \frac{1}{\tau_{\vk, \sigma}}=2 \, Z_{\vk,\sigma} \,\IM \Sigma^{\rm R}_\sigma(\vk,\omb) \left|_{\omb=\xit^\sigma_\vk} \right.,
\end{equation}
with $Z_{\vk,\sigma}$ the quasi-particle weight
\begin{equation}
Z_{\vk,\sigma} \equiv  \left[ 1- \left. \frac{\partial \, \RE  \,\Sigma^\mathrm{R}_\sigma(\vk,\omb)}{\partial \omb}\right|_{\omb=0^-} \right]^{-1}.
\end{equation}
In a standard Fermi liquid, where quasi-particles are stable, the imaginary part of the retarded self-energy is proportional to $\omb^2$ \cite{Abrikosov-1963,Nozieres-1964}, hence supporting long-lived excitations as their inverse lifetime $ 1/\tau_\vk \sim \xit_\vk^2 \ll \xit_\vk $ with $0 < Z_\vk < 1$. 

However, if the self-energy $\RE \,\Sigma^\mathrm{R}_\sigma(\vk,\omb)$ becomes singular for small $\omb$, $Z_{\vk,\sigma}$ will vanish and quasi-particles cease to exist. The system is then termed a non-Fermi liquid (NFL). Specifically, when the divergence of the derivative of the real part of the retarded self-energy is power-law with exponent between -1 and 0, one has a NFL, whereas, when the divergence is the weakest, that is logarithmic, one attains a marginal Fermi liquid. By means of the Kramers-Kronig relation between the real and imaginary parts of the retarded self-energy \cite{Gan-1993}, 
this singular behavior is reflected in the imaginary part of the retarded self-energy, which then follows a power law of the form $|\omb|^\alpha$ for small frequencies with $0< \alpha \leq 1$:   $\alpha <1$ signaling the inception of a NFL and $\alpha=1$  signifying a marginal FL. 
In the following, we examine the expression \eqref{equ:imsigret0}, 
from which the critical quantities illustrated above descend, focusing on the MSCT approach,  and  make a connection to the 3D system when relevant. We start by analyzing the pair  spectral function (or dynamical pair susceptibility) $\IM \Gamma_0^\mathrm{R}(\vQ,\Omb)$ at criticality.

\subsection{Pair spectral weight function at criticality}
\label{subsec-pair}
\begin{figure}[t]
\centering
\includegraphics[width =0.95\linewidth,trim=.4cm 0cm .5cm 0cm]{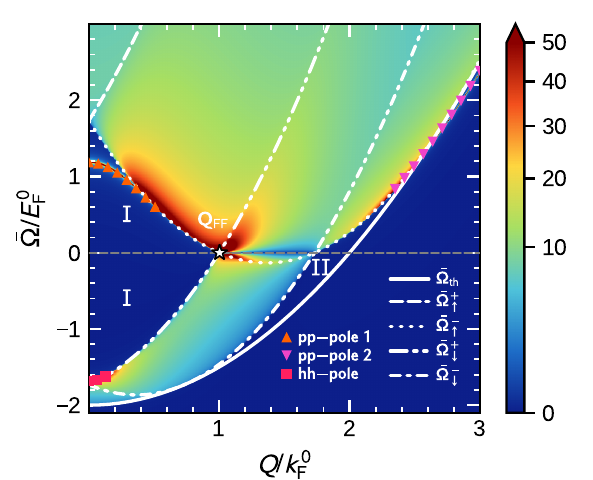}
\caption{Dimensionless pair spectral weight function $2m \, |\IM \Gamma_0^\mathrm{R}(\vQ,\Omb)|$  at coupling $\log(1/k_\mathrm{F}a_\mathrm{2D})=-0.676$ (corresponding to $\be/\eF=0.517$) and polarization $p=p_c=0.86$, within the MSCT approximation. The orange or pink triangles and magenta squares are pair collective states originating from particles and holes states, respectively. 
White lines: different threshold frequencies (see legend) defined in appendix \ref{sec:ImGammaRet}. 
Star: FFLO critical momentum $\sQfflo$.
}
    \label{fig:GammaPols}
\end{figure}

The function $ \IM \Gamma_0^\mathrm{R}(\vQ,\bar{\Omega})$ appearing in Eq.~\eqref{equ:imsigret0} is obtained by analytic continuation onto the real frequency axis of the pair propagator defined on the imaginary frequency axis in Eq.~\eqref{equ:gammamats} and reported explicitly in appendix\til\ref{sec:Thoulcrit}.

In Fig.~\ref{fig:GammaPols} we display an intensity plot of the pair spectral weight function $2m|\IM \Gamma_0^\mathrm{R}(\vQ,\bar{\Omega})|$ for a critical value of the polarization in the MSCT approximation. (We notice that since $\Gamma_0$ is a bosonic propagator, the sign of $\IM \Gamma_0^\mathrm{R}$ changes with the sign of $\bar{\Omega}$.)
One distinguishes two excitation bands
delimited by the two pairs of thresholds $\Omb_\uparrow^\pm$ and $\Omb_\downarrow^\pm$, which correspond to the majority  and minority species, respectively.
A detailed derivation  of the different energy sectors in Fig.~\ref{fig:GammaPols} and associated thresholds can be found in  appendix \ref{sec:ImGammaRet}.

In addition to the region below the vacuum threshold $\Omega_{\rm th}$ (solid line) where pairing is not allowed kinematically, particle-particle excitations  are forbidden in two additional sectors in the presence of a polarization: sector I, which is enclosed between $\Omb_\uparrow^-$ (dotted line) and $\Omb_\downarrow^+$ (dash-double-dotted line) and sector II, which  is delimited by $\Omb_\downarrow^-$ (dash-dotted line), $\Omb_\uparrow^-$ (dotted line) and the vacuum threshold $\Omega_{\rm th}$ (solid line). Within sector I, at low momenta two collective modes  originate (upward orange triangles and magenta squares), which then merge with the related continuum thresholds ($\Omb_\uparrow^-$ and $\Omb_\downarrow^+$, respectively) remaining close to them. Notably, the upper mode remains highly resonant and becomes extremely soft as its momentum approaches $\sQfflo$ (indicated by the star in Fig.~\ref{fig:GammaPols}), eventually signaling the FFLO transition. 
We interpret these modes as particle-particle (upward orange triangles) and hole-hole (magenta squares) molecular states arising merely from quantum many-body effects. We also  point out that they are also found in the 3D polarized system \cite{Urban2014}, even for coupling values below the threshold of the two-body bound state in vacuum, as opposed to the balanced system. 

Finally, at large momenta the molecular branch inherited from the balanced system
(downward pink triangles)  is recovered \cite{SVR1989}.

\subsection{Quasi-particle decay rate at criticality}
\label{sec:qpdr}

At criticality, the FFLO soft mode (star and red area in Fig. \ref{fig:GammaPols}) contributes to the intermediate state convolution in \eqref{equ:imsigret0} only if the Thouless criterion is satisfied within the range of momentum integration, that is when $\omega$ approaches zero  and when 
\begin{equation}
    |\vk^{\prime}-\vk| = Q_{\rm FF} \quad \text{and} \quad
   \xit_{k^{\prime}}^{\bar\sigma} = 0.
\end{equation}
The second relation fixes the magnitude of the intermediate state momenta $\vk^{\prime}$ to the Fermi sphere of the other species $|\vk^{\prime}| = \kLb $, whereas the first relation provides a condition on the possible magnitudes and directions of the external momentum $\vk$,
\begin{equation}
\label{equ:condfflo}
|\kLb - \sQfflo |\leq \, |\vk| \, \leq  \kLb + \sQfflo  
\end{equation}
with $\sQfflo  = \kLup - \kLdn$ 
and the angle $\gamma_k$ between $\vk$ and $\vQfflo$ uniquely identified by
\begin{equation}
   \cos\gamma_k \, =  \frac{(\kLb)^2-\sQfflosq-|\vk|^2}{2 \sQfflo |\vk|}. 
\label{equ:condfflo2}
\end{equation}
We note that if $\vk$ is on the FS of the species $\sigma$ (i.e., $|\vk|=\kL$, corresponding to the lower boundary of \eqref{equ:condfflo} if $\sigma=\downarrow$ and to the upper boundary of \eqref{equ:condfflo} if $\sigma=\uparrow$) then  $\gamma_k=0$ and $\vk$ is collinear with $\vQfflo$. For all other possible values of $\vk$ in \eqref{equ:condfflo}, $\vk$ and $\vQfflo$ form an angle $\gamma_k \neq 0$ as  given by \eqref{equ:condfflo2}.

In appendix \ref{app:imsigretonoffFS} we analytically evaluate the small-frequency behavior of the imaginary part of the retarded self-energy at criticality for both of the above cases.

For quasi-particles with momentum on the FS $|\vk|=\kLdn$ we obtain the expression (cf.~Eq.~\ref{equ:imsigret23})
\begin{equation}
\IM\,\Sigma_\downarrow^\mathrm{R}(\vkLdn,\omb)= -\frac{1}{\sqrt{3} \, 2^{4/3}}\, \frac{2m}{\kLdn \, \kLup}
\left(\frac{\kLupsq-\kLdnsq}{2m}\right)^{\frac{4}{3}} \!\!|\omb|^{\frac{2}{3}},
\label{equ:imsigret2D}
\end{equation}
illustrating a non-analytic dependence of the self-energy around $\bar{\omega}=0$. 
By Kramers-Kronig transform \cite{Gan-1993} this in turn implies the vanishing of the quasi-particle weight  and the breakdown of Fermi liquid theory \cite{Varma2002}. 
Therefore, the present system becomes a NFL when its polarization is at its critical value. An analogous result was found for a two-dimensional organic FFLO system in the weak coupling regime in \cite{Piazza2016}. 

\begin{figure}[t!]
    \includegraphics[width= 0.95\linewidth]{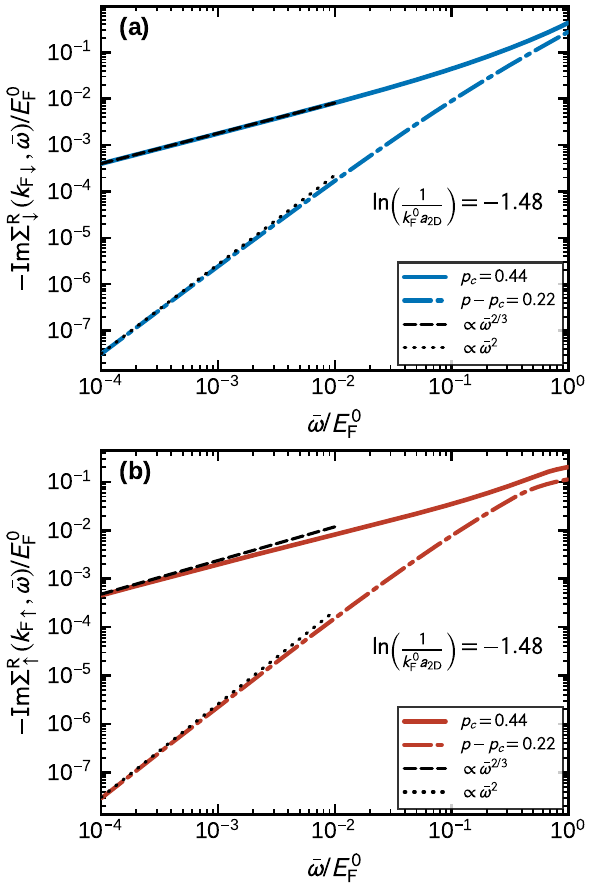}
    \caption{Imaginary part of the retarded self-energy at small frequencies for $\gtwoD=-1.48$ at critical polarization $p_c=0.22$ (solid line) and at an off-critical  polarization $p=0.44$ (dot-dashed line) for the minority (a) and majority (b) species. Dashed and dotted lines: expected analytical behaviors.}
    \label{fig:imsigretsmallw}
\end{figure}

The power law behavior of the imaginary part of the retarded self-energy at small frequencies is illustrated through a log-log plot in Fig.~\ref{fig:imsigretsmallw} at a critical and at an off-critical value of the polarization for the minority and majority  spins in panel (a) and (b), respectively. The power law $ \omega^\alpha$ is superimposed on the corresponding NFL and FL regimes with $\alpha=2/3$ and $\alpha=2$, respectively.

When the fermionic momentum $\vk$ is off the minority FS but is such that $\vkLup-\vk=\vQfflo$, the low-frequency form of the imaginary part of the retarded self-energy acquires the form (cf. Eq. \eqref{equ:imsigretoffFS})
    \begin{equation}
\IM\,\Sigma_\downarrow^\mathrm{R}(\vk,\omb)        =-\frac{1}{6 \, |\vk \times \vQfflo|}\,\frac{(\kLupsq-\kLdnsq)^2}{ \kLup \kLdn  } \,   \,|\omb|.
    \end{equation} 
The linear dependence on the frequency implies a marginal breakdown of Fermi liquid theory. 
For a marginal Fermi liquid, as originally introduced by Varma \cite{Varma2002},
the real part of the retarded self-energy behaves as $\omb \log\omb$  \cite{Gan-1993} and the quasi-particle weight vanishes in a logarithmic (weaker than power law) manner. 
    
Finally, in the remaining ranges
$|\vk| < \kLdn$ or $|\vk|>2 \kLup - \kLdn$, the imaginary part of the retarded self-energy recovers the expected FL behavior proportional to $\omb^2$ at low frequencies, as the pairing fluctuation propagator acquires an effective mass term which brings the system away from criticality and restores its FL form.

\subsection{Quasi-particle weight}
\label{sec:Zsigma}

\begin{figure}
    \centering
    \includegraphics[width=0.95\linewidth]{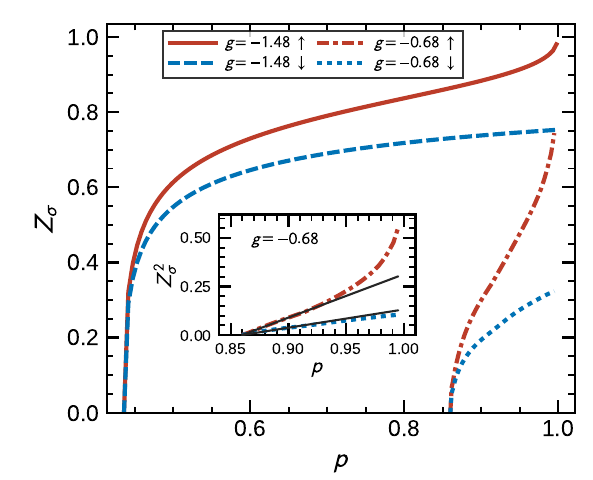}
    \caption{Quasi-particle weight $Z_\sigma$  within the MSCT scheme at  two representative coupling strengths $g=\gtwoD=-1.48, -0.68$ as a function of $p$. Inset: comparison with critical behavior $Z_\sigma \sim (p-p_c)^{1/2}$ for $g=-0.68$ (we plot $Z_\sigma^2$ for convenience).}
    \label{fig:z2d_MSCT}
\end{figure}

We now examine the evolution of the quasi-particle weight $Z_\sigma$ at the Fermi level as criticality is approached. 
This quantity is defined for quasi-particles on the FS by
\begin{equation}
    Z_\sigma \equiv  \left[ 1- \left. \frac{\partial \, \text{Re}  \,\Sigma^\mathrm{R}_\sigma(\vkL,\omb)}{\partial \omb}\right|_{\omb=0^-} \right]^{-1}
    \label{equ:qpres}
\end{equation}
(or as in Eq.\eqref{equ:Zmats} on the imaginary frequency axis) where the real part of the retarded self-energy is obtained via Kramers-Kroenig transform of Eq.~\eqref{equ:imsigret0}.
The quasi-particle weight $Z_\sigma$ satisfies $0 \leq Z_\sigma \leq 1$, which results from the property that $ \left. \frac{\partial \, \text{Re}  \,\Sigma^\mathrm{R}_\sigma(\kL,\omb)}{\partial \omb}\right|_{\omb=0^-} \leq 0$ \cite{Dupuis-2023}, as we have verified numerically in general.

At criticality, an analytic estimate of the singular behavior of the real part of the retarded self-energy at small frequencies can be obtained via the Kramers-Kroenig transform of Eq.~\eqref{equ:imsigret2D}  (see \cite{Gan-1993}) or equivalently, exploiting the uniqueness of the analytic continuation of the retarded self-energy to the upper complex  plane $\Sigma(\vkL,z)=\tilde\Sigma^0_\sigma+a\,z^{2/3}$ (with $\tilde\Sigma^0_\sigma$ defined in sec. \ref{sec:MSCT}  and $a$ a complex number). The imaginary part of $\Sigma^\mathrm{R}_\sigma(\kL,\omb)$ is an even function for small $\omb$, thus implying 
\begin{equation}
\RE\,\Sigma_\sigma^\mathrm{R}(\vkL,\omb)= \tilde\Sigma^0_\sigma+\sqrt{3} \, \sgn(\omb)  \,\IM\,\Sigma_\sigma^\mathrm{R}(\vkL,\omb),
\label{resigma}
\end{equation}
which leads to $ \partial \, \text{Re}  \,\Sigma^\mathrm{R}_\sigma(\kL,\omb)/\partial \omb\to -\infty$ when $\omb\to 0$ and thus a vanishing quasi-particle weight at criticality.

We report $Z_\sigma$ in Fig. \ref{fig:z2d_MSCT} as a function of polarization at two representative coupling strengths $\gtwoD=-1.48$ and $\gtwoD=-0.68$ for both species. As the polarization approaches its critical value, the quasi-particle weight is seen to vanish.

In particular, we have numerically verified that $Z_\sigma \sim (p-p_c)^{1/2}$ when approaching the critical polarization (see the inset). 
In Ref.~\cite{Senthil2008} it has been quite generally argued that, at a quantum critical point associated with a vanishing quasi-particle weight, one should have $Z_\sigma \sim (p-p_c)^{\nu(z_f-d_\alpha)}$,
where $\nu$ is the correlation length critical exponent, $z_f$ is the fermionic dynamical critical exponent, and $d_{\alpha}$ is the scaling exponent of  the fermionic spectral weight function for $\vk \simeq \vkL $ and $\omb \simeq 0$ \cite{Senthil2008,Pini2023}:
\begin{equation}
A_\sigma(\vk,\omb)\sim \frac{c_{0 \sigma}}{ |\omb|^{d_\alpha / z_f } }\, F_0\left(c_{1\sigma} \frac{\omb}{||\vk|-\kL|^{z_f} }\right),
\label{eq:Senthil}
\end{equation}
where $F_0$ is a universal scaling function and $c_{0 \sigma}, c_{1 \sigma}$ are
nonuniversal coefficients.

We will see in sec.~\ref{sec:dynexpF}
that  $z_f=3$. Expressions 
\eqref{equ:imsigret2D} and \eqref{resigma} yield
 $A(\vkL,\omb) \sim \bar{\omega}^{-2/3}$ and one obtains $d_\alpha=2$.
In appendix \ref{sec:appA2} we derive the value $1/2$ for the correlation length exponent $\nu$, so finally we obtain
the behavior $Z_\sigma \sim (p-p_c)^{1/2}$ that we have verified numerically.

\subsection{Momentum distribution function $n_\sigma(k)$: position and closure of the Fermi step}

\begin{figure}[t!]
    \centering
    \includegraphics[width=0.95\linewidth]{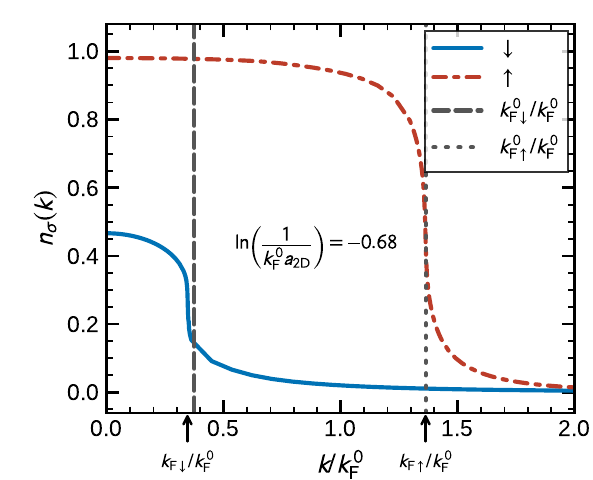}
    \caption{2D momentum distribution function $n_\sigma(k)$ within the MSCT approach  at critical polarization $(p_c=0.860)$ for $\gtwoD=-0.68$.}
    \label{fig:nk2d} 
\end{figure}

The above features are reflected in the momentum distribution functions $n_\sigma(k)$ shown in Fig.~\ref{fig:nk2d} at criticality for a representative coupling within the MSCT scheme. The Fermi step is  seen to close for both species, confirming the NFL behavior discussed above. The positions of the steps (determined by the Fermi momentum $\kL$) are identified by the respective vertical arrows, while the Fermi momenta of the corresponding non-interacting system ($\kFs$) are indicated by a vertical line (dotted for spin up, dashed for spin down). The discrepancy between the interacting and non-interacting Fermi momenta is a measure of the residual deviation of the MSCT approach from the Luttinger sum rule \cite{Luttinger1960,Luttinger2}. 
We find a negligible deviation for the up spins and a small one for the down spins.

\begin{figure}[t!]
    \centering
    \includegraphics[width=0.95\linewidth]{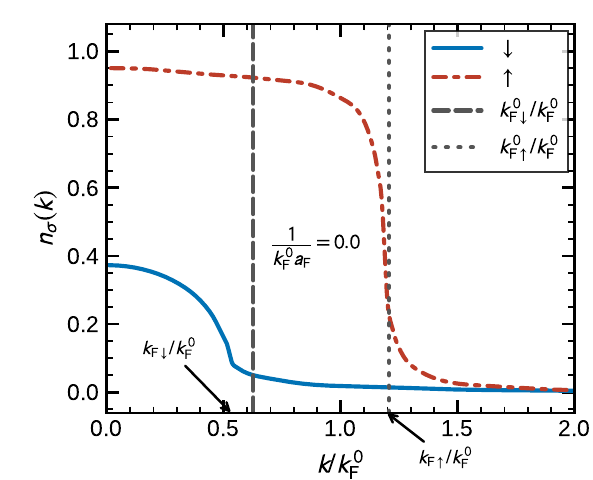}
    \caption{3D momentum distribution function $n_\sigma(k)$ within the MSCT approach  at critical polarization $(p_c=0.754)$ for $1/(\kF a_\mathrm{F})=0$.}
    \label{fig:nk3d} 
\end{figure}

It is interesting to make a comparison with the momentum distribution of the 3D system reported in Fig.~\ref{fig:nk3d}.
As in 2D, the Fermi step disappears for both species, but a somewhat larger discrepancy between the interacting and non-interacting Fermi momenta is observed in 3D, especially for the minority species. These differences are in line with our previous discussion of the critical polarizations in Figs.~\ref{fig:phasediag} and \ref{fig:phasediag3D}, and the role of the Luttinger theorem in determining the  deviation between the mean-field curves and the MSCT ones.

\section{Dynamical critical exponents}

The critical behaviors of the spatial ($\xi$) and temporal ($\xi_\tau$) relaxation scales of (bosonic) critical fluctuations are connected by the dynamical exponent $z_b$ through $\xi_\tau \propto \xi^{z_b}$ \cite{Hertz-1976,HH-1977}. When critical fluctuations have a  propagating nature (as e.g., for sound-like excitations) 
the dynamical exponent is generally $z_b=1$, whereas $z_b=2$ or $3$ characterizes diffusive critical regimes \cite{HH-1977}.

In the following, we analyze the dynamical critical exponents of the 2D bosonic and fermionic excitations of the FFLO transition within the MSCT approximation. This scheme is essentially equivalent to what is termed non-self-consistent random phase approximation (RPA) in the context of quantum phase transitions \cite{Holder-2015-2}. 

In addition, extending the analysis of the 3D system in \cite{Pini2023}, we go beyond the MSCT (RPA) approach and demonstrate that the vertex corrections to the Hertz-Millis (HM) theory of quantum phase transitions \cite{Hertz-1976,Millis-1993} in the pairing channel are irrelevant to all orders, thus placing the 3D FFLO transition in the mean-field universality class.

\subsection{Bosonic dynamical exponent $z_b$ in 2D}

\label{sec:dynexpB}

The scaling behavior of the pairing fluctuations at criticality can be obtained by evaluating the critical bosonic propagator in the low energy limit and for pair momenta close to the FFLO critical line, $|\vq|=|\vQ-\vQfflo| \ll \sQfflo$.

The critical bosonic propagator  $\Gamma_0(\vQfflo+\vq,i\Omega)$ is shown in appendix \ref{sec:gammalowE} to take the following form,
\begin{multline}
-\frac{4\pi}{m} \Gamma_0(\vQfflo+\vq,i\Omega)^{-1}= \alpha\, \frac{\Omega }{\sqrt{ \frac{q_t^2}{4\sQfflosq}+\frac{q_r}{2\sQfflo} } }  \\ + \beta \, \left(\frac{q_t^2}{4\sQfflosq}+\frac{q_r}{2\sQfflo}\right),
\label{equ:gamscal}
\end{multline}
having defined $\alpha \equiv (m \, \vff \, \sqrt{\vLup \vLdn} )^{-1}$
and $\beta \equiv \vLup \vLdn / v^2$, with
$v=(\vLup+\vLdn)/2$, $\vL=\kL/m$ and $ \vff=\sQfflo/m $.
The momenta $q_r$ and $q_t$ are the radial and  transverse components of the small momentum $\vq$ with respect to a fixed $\vQfflo$,  which can be arbitrarily chosen along the circle $|\vQ|=\sQfflo$.

We restrict to momenta such that 
\begin{equation}
e_Q^\sigma \approx v_{\sigma} \left(\frac{q_t^2}{2{\sQfflo}}+q_r\right)>0,
\end{equation}
since these are the dominant modes contributing to the fermionic dynamics at criticality in Eq.~\eqref{equ:imsigret0} (see discussion about the critical exponent $\nu$ in appendix \ref{sec:Thoulcrit}).

Expression \eqref{equ:gamscal} is reminiscent of the dynamic susceptibility of a number of 2D quantum itinerant systems at criticality: i) a 2D FL close to a Pomeranchuk instability \cite{Oganesyan-2001,Varma2002,Metzner2003,Metzner2006,Wolfle2007} (with partial wave component $l=0$ denoting the itinerant ferromagnet and $l=2$ the nematic ordering), ii) the transition to the 2D charge density wave order \cite{Holder2014}, and, more generally, iii)  a 2D FL coupled to a U(1) gauge field \cite{Lee-1989,Gan-1993,Metlitski-2010,Holder-2015}.  
Scale invariance 
under the transformations
\begin{equation}
\label{equ:scaltrsB}
    q_t \rightarrow \lambda \, q_t, \qquad 
    q_r \rightarrow \lambda^2 \, q_r, \qquad \Omega \rightarrow \lambda^{z_b} \, \Omega,
\end{equation}
implies the dynamical  exponent $z_b=3$, in line with the predictions of the RPA (which corresponds to the gaussian fixed point of the associated Landau-Ginzburg-Wilson functional) for the aforementioned quantum critical systems \cite{Gan-1993,Oganesyan-2001,Metlitski-2010, Holder-2015}.
According to the standard theory of quantum phase transition in itinerant electron systems introduced by Hertz and Millis \cite{Hertz-1976,Millis-1993}, the effective dimension reflecting the mixing of static and dynamic correlations at zero temperature is $d+z_b=5$ for $d=2$ and $z_b=3$, which is greater than the upper critical dimension $d_c^+=4$. This makes the RPA result stable at all levels of renormalization. 

However, dimensionality $d$ plays a crucial role in the assumptions of the theory (originally illustrated by  Hertz  for the itinerant ferromagnet in $d=3$ \cite{Hertz-1976}). Belitz, Kirkpatrick and  Vojta \cite{BKV1997} have shown that, once vertex corrections are included beyond RPA, the static triplet susceptibility of the ferromagnet is non-analytic and non-local in $1 < d < 3$ ($\chi(\vQ,0)^{-1} \sim |\vQ|^{d-1}$) as well as in $d=3$ (with $\chi(\vQ,0)^{-1} \sim \vQ^2 \log|\vQ|$)  \cite{Belitz-1998,Chubukov-2003}. This implies that the gradient expansion of the Landau-Ginzburg-Wilson functional employed by Hertz breaks down. Subsequently, Chubukov has reassessed the problem beyond the static approximation and confirmed the breakdown of the HM theory for critical ferromagnetic fluctuations \cite{Chubukov-2004-FM,Chubukov-2006}. 
In general, whenever gapless bosonic fluctuations are coupled to gapless fermionic excitations (as in itinerant systems) the HM theory may become inadequate in 3D and definitively is in 2D \cite{Sachdev2011}. 
Seminal results were also obtained for the itinerant antiferromagnet by Chubukov in \cite{Chubukov-2004-AFM}, whereby the prediction of the HM theory was   proven to continue to hold in 3D but to break down in 2D.

A number of approaches have been put forward to go beyond the HM paradigm and treat fermionic and bosonic degrees of freedom on the same footing.
Concerning the present problem of the FFLO criticality in 2D and the identification of the related upper critical dimension, recently the authors of \cite{Pimenov_2018} have adopted the dimensional regularization scheme developed by Lee {\it et al.} for Fermi liquid instabilities with zero ordering wave-vector and parabolic FS's (as in the 2D Ising-nematic transition) \cite{Lee-2013,Mandal-2015}. They could identify the upper critical dimension $d_c^+=5/2$ by relying on the equivalence between the topologies of the FS's involved in the Ising nematic transition and in the 2D FFLO transition. In addition, they found a scaling behavior of the bosonic propagator that is consistent with the frequency dependent term $\sim \Omega/q$ (as in Eq.\eqref{equ:gamscal}).   
However, the outcome of this approach is strongly dependent on the FS topology being one-dimensional and parabolic (as in the 2D Ising-nematic instability treated by Lee \cite{Lee-2015}). As a consequence, it cannot be applied to the case of the 3D FFLO transition in which the FS is two-dimensional and locally flat. This case is discussed just below.

\subsection{Bosonic dynamical exponent and vertex corrections in 3D}
\label{sec:boszvert}

In 3D, the fluctuation propagator (or pair susceptibility) has the form
\cite{Perali2002,Pini2023}  
\begin{equation}
    \Gamma_0(\vQ,i\Omega)=\frac{1}{a+b\,(|\vQ|-\sQfflo)^2+(d_1+i\,d_2 \,\text{sgn}[\Omega])\,i\,\Omega}
    \label{equ:gammats3D}
\end{equation}
with $a=0$ at criticality, from which it is straightforward to read off the mean-field critical exponents for space correlations $\nu=1/2$ and for time correlations $z_b=2$ \cite{Pini2023}.

Espression \eqref{equ:gammats3D} is reminiscent of the RPA susceptibility of the itinerant antiferromagnet (IAFM) in 3D heavy-fermion systems \cite{Varma2002,Steglich-2006}, even though in the latter case the dispersion of the spin density wave  $\propto (\vQ-\vQ_\mathrm{AFM})^2$ is non-isotropic owing to the underlying lattice \cite{Fawcett-1988}. The 3D IAFM shares the same mean-field critical exponents $\nu$ and $z$ with the 3D FFLO transition and its vertex corrections to the RPA susceptibility have been shown in \cite{Chubukov-2004-AFM} to be irrelevant at all loop orders, thus demonstrating that the 3D IAFM belongs to the mean-field universality class.

According to the HM theory \cite{Hertz-1976,Millis-1993} the 3D FFLO transition should have a gaussian fixed point as well, since its effective dimension $d+z_b=5$ is greater than the upper critical dimension $d_c^+=4$. 
However, Hertz's assumption of well-behaved vertex corrections to the bosonic susceptibility at small $\vQ $ and $\Omega$  \cite{Hertz-1974}, has been shown to be incorrect for most itinerant systems \cite{BKV1997,Belitz-1998,Chubukov-2003,Chubukov-2004-AFM,Chubukov-2004-FM,Chubukov-2006},
as the gapless excitations are not only bosonic but also fermionic.  The latter may generate non-analytic Landau coefficients (namely singular vertex corrections)  after being integrated out in the process of obtaining an effective free energy expansion in the order parameter only \cite{Abanov2003}.

\begin{figure}[t!]
\includegraphics[width=0.45\textwidth]{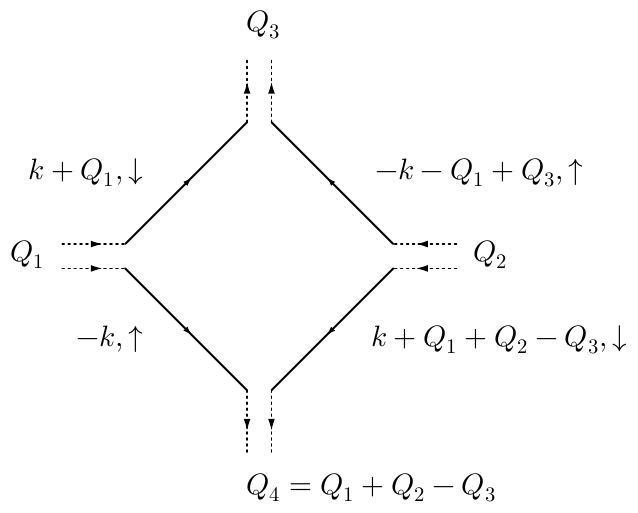}
\caption{4-point boson vertex of the Hertz-Millis functional within the particle-particle channel, corresponding to the effective two-boson interaction $u_2$ of Ref. \cite{Strinati2000}.}
\label{fig:b4diag} 
\end{figure}

Motivated by the analogy between the FFLO and IAFM susceptibilities, we now turn to a three-dimensional lattice geometry. This setting enables us to assess the robustness of our RPA-level results upon including boson–boson interactions, closely following the approach developed in Ref.~\cite{Chubukov-2004-AFM}.
In appendix  \ref{app:4bosvert}  we explicitly evaluate the first vertex correction to the RPA pair susceptibility \eqref{equ:gammats3D}, namely the 4-point boson vertex $u_2$ depicted in Fig.~\ref{fig:b4diag} (we follow the same notation of \cite{Pistolesi-1996}, with $u_n$ indicating the $2n$-point bosonic vertex). 
We show that the coefficient $u_2(Q_1,Q_2,Q_3,Q_4)$, where $Q_i\equiv(\vQ_i,i\Omega_i)$ are four-momenta,
does not have a uniquely defined limit for $\vQ_i \to \vQfflo$ and $\Omega_i \to 0$. We also show that  $u_2$  becomes singular if the limit of small frequencies is approached at $\vQ_i=\vQfflo$. 
This, in turn, produces pairing  correlations that decay algebraically  in time, thus breaching the requisite of locality in time (and space) which underpins the validity of the Landau-Ginzburg-Wilson expansion performed by Hertz at zero temperature \cite{Hertz-1976}.

We find  that the scaling dimension of this vertex in 3D is $[u_2]=z_b-2$ with $z_b>1$ (rather than $[u_2]=0$ of the HM theory), and that its running coupling constant $g_4$ scales as $[g_4]=3-2z_b$ (rather than $[g_4]=4-d-z_b$ with $d=3$ of the HM theory \cite{Hertz-1976}). In the present system $z_b=2$, hence the 4-point vertex is found to be irrelevant and the HM prediction is accidentally confirmed. 
    The same conclusion can be drawn for the 3D Ising nematic transition \cite{Oganesyan-2001}, but the notable exception in 3D is represented by the itinerant ferromagnet \cite{BKV1997,Chubukov-2004-FM}, where the 4-point vertex correction generates non-analytic behavior even in the static susceptibility for vanishing momentum \cite{Chubukov-2004-FM}. In 2D, we recall that the HM expansion consistently breaks down for itinerant systems because it produces marginal or relevant vertex corrections at all loop orders  \cite{Sachdev2011}. 

Higher order vertex corrections can be 
computed following the integration scheme adopted in appendix \ref{app:4bosvert}. 
By aligning two of the three momentum integrations with the majority and minority Fermi velocities, these integrals become separable and effectively one-dimensional.
The remaining integration gives an overall factor proportional to the  momentum cutoff in the transverse direction.
By induction, one can write for the vertex of order $n$ \cite{Strinati2000}, 
\begin{equation}
    u_{n} \propto \frac{|\Omega|}{(i\Omega-\vLup q)^{n-1}(i\Omega-\vLdn q)^{n-1}},
\end{equation}
where the quantity $q=|\vQ-\vQfflo|$ and we are using the same compact notation
as in \cite{Chubukov-2004-AFM}. 
The vertex of order $n$ thus scales as $[u_n]=z_b-2\,(n-1)$ and enters the $n$-th contribution to the boson effective action \cite{Pistolesi-1996,Strinati2000} through the expression
\begin{equation}
   g_{2n} \, \int (d^4Q_i)^{(2n-1)} \; u_n \; (\Psi_{Q_i})^{2n},
\end{equation}
where $g_{2n}$ is the running coupling constant subject to the renormalization flow.
This quantity scales as 
\begin{align}
    [g_{2n}]&=-\{\underbrace{(3+z_b)\,(2n-1)}_{\textstyle[(d^3Q\,d\Omega)^{2n-1}]}+\underbrace{z_b-2\,(n-1)}_{\textstyle[u_n]}+\underbrace{n\,(-5-z_b)}_{\textstyle[\Psi_Q^{2n}]} \} \nonumber\\   
    & = n+1-n\,z_b,   
\end{align}
which means that it is irrelevant for all $n > 1$  since  in the present case $z_b=2$.

We can therefore conclude that the critical behavior at the FFLO transition on a 3D lattice is reproduced by the gaussian fixed point of Hertz's theory, and the resulting critical exponents $\nu=1/2$ and $z_b=2$ belong to the mean-field universality class. 
This class is shared with the 3D IAFM, which was shown in \cite{Chubukov-2004-AFM} to have mean-field critical exponents as well.

\subsection{2D vs 3D nesting of FS's}
\label{sec:FSnesting}
A remarkable difference between the 2D and 3D expressions of the pairing fluctuation propagator at criticality (Eqs.~\eqref{equ:gamscal} and \eqref{equ:gammats3D}, respectively) emerges in the term proportional to the frequency (the so-called damping term \cite{Abanov2003,Vojta2007}).
In 3D, the damping term is $\sim \Omega$, while in 2D it is $\sim \Omega/q$ (where we approximate $q_t \approx q$ since the longitudinal momentum can be assumed much smaller than the tangential one \cite{Metlitski-2010,Lee-2008,POLCHINSKI-1994}).

\begin{figure}
    \includegraphics[width=0.3\textwidth]{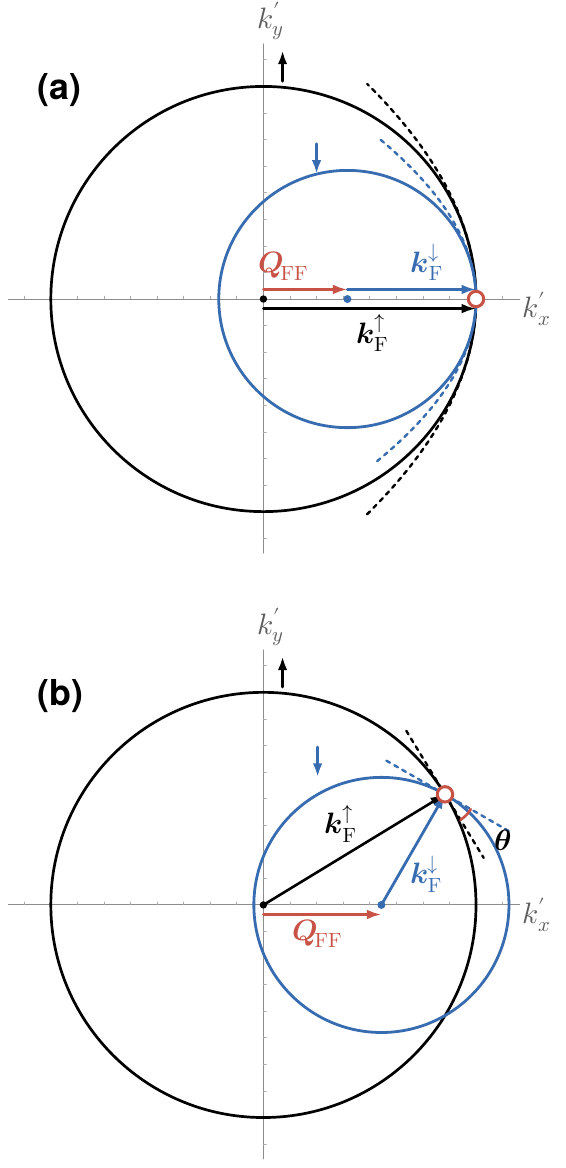}
    \caption{Geometrical configuration of majority and minority FS's whith pairing momentum  (a) $\sQfflo=\kLup-\kLdn$ and (b) $\sQfflo >\kLup-\kLdn$ for a given directional choice of $\vQfflo$. The configuration in (a) occurs in 2D and results in a local parabolic approximation (dashed lines,  Eq.~\eqref{eq:newfs}) at the matching point of the two FS's (red circle). The configuration in (b) occurs in 3D (the cut with $k_z^\prime=0$ is reported in this case) and results in a flat approximation (dashed lines, Eq.~\eqref{eq:newfs2}) at the matching point (red circle). The angle $\theta$ determines the local nesting between the two FS's and is equal to 0 for the configuration in panel (a) (locally perfect nesting).} 
\label{fig:FSgeom}
\end{figure}
The origin of this difference can be understood by looking at the geometry of
the region of momenta $\mathbf{k^\prime}$ that is dominant in determining the result of convolution \eqref{equ:gammaseries} at criticality (with $G_{0\sigma}$ replaced by $\tilde{G}_{0\sigma}$ within MSCT).

This region is defined by requiring $\mathbf{k^\prime}$ and $\vQ-\mathbf{k^\prime}$ to be on the FS of the minority and majority species, respectively, with $|\vQ|\simeq\sQfflo$. This condition on $\mathbf{k^\prime}$ can be satisfied only if $|\vkLup-\vkLdn|=Q_{\rm FF}$. This relation defines a manifold—such as a point or a circle—where the two FS's match and pairing is optimal.
In 2D, where $Q_{\rm FF}=\kLup-\kLdn$, the two FS's are tangent at the matching point (see Fig.\til\ref{fig:FSgeom}(a)).
As a result, the local geometry can be approximated by a parabola (cf Eq.~\eqref{eq:newfs} in appendix \ref{app:imsigretonoffFS}), similarly to the 2D Ising nematic transition \cite{Oganesyan-2001,Metlitski-2010} and the 2D charge density-wave order \cite{Holder2014} (but differently from the itinerant antiferromagnet \cite{Sachdev2011}). 
When singular fermionic propagators with parabolic dispersion of the form
\begin{equation}
    \tilde{G}_{0\sigma}(\vk,i\omega)^{-1}=i\omega-\frac{\left[ (\vk-\vkL) \times \vvL\right]^2}{2m \, |\vvL|^2}-\vvL \cdot ( \vk-\vkL)
    \label{equ:softGf}
\end{equation}
(see again Eq.~\eqref{eq:newfs}) are inserted into the convolution \eqref{equ:gammaseries} of the bosonic propagator, the resulting damping term is of the type $\Omega/q$ (as shown explicitly in \cite{Metlitski-2010}). 

In contrast, for the FFLO transition in 3D, one has $Q_{\rm FF} > \kLup-\kLdn$ (see Fig.~\ref{fig:QffloVScoupl}), implying that the two FS's are crossing rather than tangent at the matching point (given the spherical symmetry in 3D, this point is actually a circle). 
As a consequence, the local geometry at this point is flat (see Fig.~\ref{fig:FSgeom}(b)) and  the resulting soft fermionic propagator
has the form
\begin{equation}
    \tilde{G}_{0\sigma}(\vk,\omega)^{-1}=i\omega-\vvL \cdot ( \vk-\vkL).
\label{equ:got3D}
\end{equation}
One can show that, after inserting the above propagator in the 3D counterpart of the convolution \eqref{equ:gammaseries},  a damping term of the form $\sim \Omega$ is obtained (see \cite{Sachdev2011} for details).

The non-collinearity between the three momenta $\vkLup,\vkLdn$ and $\vQfflo$ in  Fig.\til\ref{fig:FSgeom}(b)
is reflected by a non-zero value of the  nesting parameter $\tan(\theta)$, measuring the angle between the two Fermi velocities at the matching point in Fig.~\ref{fig:FSgeom}(b).

We note that, because of  rotational invariance of the present continuum case, the direction of the vector $\vQfflo$ is arbitrary, thus allowing every point on the FS's to act as a matching point. On a lattice, instead, rotational invariance is broken, restricting $\vQfflo$ to a finite number of values connected by the discrete symmetries of the lattice. The corresponding discrete matching points on the FS's are usually indicated as ``hot spots" or ``hot manifolds" \cite{Chubukov-2004-AFM,Sachdev2011}. In the present case the entire FS's can be considered as ``hot" when the rotational invariance of $\vQfflo$ is taken into account.

It is thus interesting to draw a parallel again with the IAFM at the level of the RPA approach. 
A flat geometry of the FS at the matching point is known to occur both in the 3D and in the 2D IAFM \cite{Chubukov-2004-AFM,Sachdev2011} resulting in the expected damping term $\sim \Omega$, as illustrated in \cite{Metlitski-2010-II,Sachdev2011}.
While in 3D the IAFM has stable mean-field critical exponents \cite{Chubukov-2004-AFM}, RG studies in 2D \cite{Strack-2016,Lee-2017,Lee-2017-2} have shown that the nesting parameter $\tan(\theta)$ is renormalized to zero and the related dynamical exponent $z_b$ changes from 2 to 1. These RG results for $\tan(\theta)$ and $z_b$ have been quantitatively refined by a recent QMC study in Ref.\til\cite{Lindsey2023}.
Unlike the 2D IAFM, the 2D FFLO instability exhibits an entirely different FS geometry ($\theta=0$ in Fig.~\ref{fig:FSgeom}(a)) with respect to its 3D counterpart. A renormalization group analysis at the  one-loop order~\cite{Pimenov_2018} confirmed the RPA value $z_b=3$~\cite{Piazza2016} of the  bosonic dynamical critical exponent.

Finally, we mention that a geometric reasoning analogous to that illustrated in Fig.\til\ref{fig:FSgeom} was used in Ref.~\cite{Norman-2004} in relation to  the 3D IAFM Chromium.

\subsection{Fermionic dynamical exponent}
\label{sec:dynexpF}

In a critical itinerant fermionic system, soft modes appear not only as collective (bosonic) fluctuations but also as gapless fermionic excitations \cite{BK-2016}. We are therefore interested in the low energy form of the fermionic propagator at criticality and we consider its  expansion for fermionic momenta close to the FS and small frequencies. 

\subsubsection{2D}

By analytically continuing the retarded self-energy at criticality [Eqs.~\eqref{equ:imsigret2D} and \eqref{resigma}] to the positive imaginary frequency axis, we obtain the following expression for the dressed fermionic propagator  
\begin{align}
    \tilde G_\sigma(\vkL+\vk,i\omega)^{-1}&=-i\, \gamma \, |\omega|^{2/3} - \frac{{k_y}^2}{2m} -\vL k_x.
\label{equ:Gfcrit}
\end{align}
Here, we adopt the conventional low-energy formulation of Fermi liquid theory \cite{Sachdev2011}, where momenta are measured from the FS ($\vk \to \vkL+\vk$) and decomposed  into  radial ($x$) and transverse ($y$) components, with $\vkL$ parallel to the local $x$ axis. Furthermore, $\vL=\kL/m$, the coefficient 
\begin{equation}
\gamma=\frac{1}{\sqrt{3} \, 2^{1/3}}\, \frac{2m}{\kLdn \, \kLup}
\left(\frac{\kLupsq-\kLdnsq}{2m}\right)^{4/3},
\end{equation}
and the propagator $\tilde G_\sigma$ indicates the MSCT bare propagator corrected by the inclusion of the dominant frequency-dependent part of the self-energy.

Applying to Eq.~\eqref{equ:Gfcrit} the  scaling transformations (as for the bosonic case of Eq.~\eqref{equ:scaltrsB}),
\begin{equation}
\label{equ:scaltrsF}
    k_y \rightarrow \lambda \, k_y, \qquad 
    k_x \rightarrow \lambda^2 \, k_x, \qquad \omega \rightarrow \lambda^{z_f} \, \omega,
\end{equation}
we find a fermionic dynamic exponent $z_f=3$ (which thus coincides in value with the bosonic dynamic critical exponent).
An equivalent exponent was found in \cite{Piazza2016}  with a reported value of $3/2$ owing to a different choice of the principal scaling coordinate -- which was the radial one ($x$ axis) rather than the transverse one ($y$ axis) used in this work \cite{Metlitski-2010,Senthil2008}. 

We note that the self-energy exponent $2/3$ signaling the NFL behavior in Eq.\til\eqref{equ:Gfcrit} stems from the specific geometry of the loci of points of singularity (soft modes) of the bosonic and fermionic propagators appearing in the convolution of the self-energy \eqref{equ:se}. As illustrated in Fig.~\ref{fig:hotspot} of appendix \ref{app:imsigretonoffFS}, the fermionic soft modes (the locally parabolic FS as implied by Eq.\til\eqref{equ:Gfcrit}), and the bosonic soft modes (the red circle locating the singularities of the pairing propagator) are tangent to each other. This configuration, combined with an $\Omega/q$  bosonic damping term, generates a fermionic damping term scaling as $\sim \omega^{2/3}$.

The same behavior is found across several 2D quantum critical metals at the RPA level, as in the 2D Ising nematic transition \cite{Oganesyan-2001,Metlitski-2010}, the itinerant ferromagnet \cite{Chubukov-2004-FM} and the charge density wave transition \cite{Holder2014} (the nematic and itinerant ferromagnet transition are also known as Pomeranchuk instabilities in the spin channel with $l=2$ and $l=0$, respectively). These critical systems share with the FFLO transition the same form of the bosonic damping term $\Omega/q$ (and therefore the same $z_b$) and a local parabolic geometry of the FS, resulting in the same frequency dependence of the fermionic self-energy on the FS ($\sim \omega^{2/3}$). We note that the above analogies hold regardless of the fact that the ordering wave vector is zero (Pomeranchuk instabilities) rather than finite (charge density wave). However, we stress that crucial physical differences between these systems emerge  when going beyond the one-loop level of renormalization to include vertex corrections \cite{Metlitski-2010,Chubukov-2006}. 
\begin{table*}[t!]
    \centering
\begin{tabular}{|cc|c|c|c|c|} \toprule
                          &       & Hertz-Millis &  Boson Damping   & Fermion Damping    & Ref. \\ \midrule
\multirow{2}{*}{Itinerant FM, $Q_0=0$} 
                          &  2D  &     No  & $\Om/q$ & $\om^{2/3}$   &   \cite{Chubukov-2004-FM,Chubukov-2006}   \\
                          &  3D  &     No  & $\Om/q$ & $\om^{1}$    &   \cite{Vojta2007}   \\  \midrule
\multirow{2}{*}{Ising-nematic, $Q_0 =0$}    
                          &  2D  &     No & $\Om/q$ & $\om^{2/3}$    &   \cite{Metlitski-2010}  \\
                          &  3D  &     Yes & $\Om/q$ & $\om^{1}$      &   \cite{Oganesyan-2001}  \\ \midrule
\multirow{1}{*}{CDW, $Q_0\neq0$} 
                          &  2D Discrete &     No   & $\Om/q$   & $\om^{2/3}$   &   \cite{Holder2014,Mandal2024}  \\ \midrule
\multirow{2}{*}{Itinerant AFM, $Q_0\neq0$} 
                          &  2D Discrete &     No   & $\Om$   & $\om^{1/2}$   &   \cite{Metlitski-2010-II,Sachdev2011}  \\
                          &  3D Discrete &     Yes & $\Om$   & $\om^{1}$     &   \cite{Chubukov-2004-AFM,Vojta2007}   \\ \midrule
\multirow{2}{*}{FFLO, $Q_0\neq0$}     
                          &  2D  Continuum &  No   & $\Om/q$  & $\om^{2/3}$  &   this work    \\
                          &  3D  Continuum &  Yes  & $\Om$   & $\om^{1/2}$    &   this work $\&$ \cite{Pini2023}   \\ \midrule
\end{tabular}
\caption{Results on various examples of critical Fermi liquids (in a continuum or discrete geometry) at the RPA level of approximation.  For discrete geometries, we consider only ordering wave-vectors $\mathbf{Q_0}$ that are commensurate (incommensurate) with the lattice 
for the IAFM (CDW). For each system (first column)  we report whether the standard Hertz and Millis theory holds (second column), the form the damping term 
of the bosonic (third column) and fermionic (fourth column) soft modes, and relevant references (last column).}
\label{tab:systsumm}
\end{table*}

\subsubsection{3D}
A similar reasoning can be applied to the 3D FFLO transition by taking into account that the bosonic damping term is now $\sim \Omega$, as discussed in sec.~\ref{sec:dynexpB}. Moreover, the locus of points in momentum space (red circle) hosting the soft modes of the pairing propagator is now secant rather than tangent to the FS of either species, 
generating a configuration analogous to that shown in Fig.\til\ref{fig:hotspotnoncoll} 
of appendix \ref{app:imsigretonoffFS}
for the 2D system. 
As previously discussed in sec.~\ref{sec:FSnesting}, the geometrical condition 
$|\vkLup-\vkLdn|=Q_{\rm FF}$ with $Q_{\rm FF} > \kLup-\kLdn$
implies that the three momenta $\vQfflo,\vkLup,\vkLdn$ are non-collinear.
Consequently, this geometric configuration dictates that the aforementioned locus of points must intersect, rather than touch, the FS of either species.

In appendix \ref{app:imsigret3D} we perform an analytical estimate of the imaginary part of the retarded self-energy at low frequency at criticality. From this calculation we obtain that for $\omega \to 0^+$ the dressed fermionic propagator acquires the low energy form 
\begin{equation}
    \tilde{G}_{\sigma}(\vkL+\vk,i\omega)^{-1}=(B_\sigma-i \, C_\sigma)\,\omega^{1/2}-\vvL \cdot  \vk,
    \label{equ:dressedG}
\end{equation}
with $B_\sigma,C_\sigma$ positive constants that are obtained from the  analytic continuation of expression \eqref{equ:imsigR3d} to the positive imaginary frequency axis. This confirms what found numerically in \cite{Pini2023} at the level of self-consistent $t$-matrix.

Applying the scaling transformations 
\begin{equation}
    \vk \rightarrow \lambda \, \vk, \qquad 
    \omega \rightarrow \lambda^{z_f} \, \omega,
\end{equation}
we obtain the fermionic dynamic critical exponent $z_f=2$ in 3D (which also in this case has the same value as the bosonic dynamic critical exponent $z_b$).

As a final remark, we note that 
for the 3D IAFM within the RPA, the  fermionic self-energy exhibits a marginal FL behavior ($\sim\omega$) \cite{Vojta2007}, rather than the NFL behavior found here for the 3D FFLO transition in the continuum.
This difference originates from their  corresponding ordering wave vectors, which have a preferred direction for the IAFM on a lattice and no preferred direction for the FFLO transition in the continuum \cite{Sachdev2011}.
 
\subsection{Comparison with other quantum critical metals}

In summary, Table \ref{tab:systsumm}  compares our results for the bosonic and fermionic damping, as well as the validity of the HM theory, against several  widely studied quantum-critical itinerant systems at the RPA level (which is equivalent to the present MSCT approach in the particle-particle channel). The bosonic damping term appears in the fluctuation propagator computed at $\vQ-\mathbf{Q_0}=\vq$, where $\mathbf{Q_0}$ is the ordering wave-vector. The fermionic damping term appears in the fermionic propagator computed at $\vk=\vkL$  \cite{Vojta2007}.

The predictions of the Hertz-Millis theory appear essentially applicable in 3D for most quantum systems, with the exception of the itinerant ferromagnet. Conversely, HM theory breaks down in 2D, although (at the RPA level of approximation) it still predicts the correct (non- or marginal-) Fermi liquid behaviors in the fermionic excitation sector.

Note also that the damping form $\Om/q$ is always  associated with a locally parabolic behavior of the FS at the matching point. This is always the case when the ordering wave-vector $Q_0=0$, whereas the damping form $\Om$ arises from the possibility of linearizing the FS at the matching point \cite{Sachdev2011}. This generally happens when $Q_0 \neq 0$.

\section{Conclusions}

We have investigated a 2D polarized Fermi gas at zero temperature from the normal phase down to the superfluid phase transition across the full range of coupling strengths and polarizations within a diagrammatic $t$-matrix approach. We have shown that some self-consistence in the fermionic propagators is necessary to obtain a sensible critical polarization and overcome the large violations of the Luttinger theorem that occur in the non-self-consistent approach. To this end, we have implemented a minimal scheme  that incorporates self-consistency in an approximate way (MSCT approach). 

Within this scheme, we have analyzed the quasi-particle decay rate on the two FS's and found, both numerically and analytically, a NFL behavior of the imaginary part of the retarded self-energy, which in turn implies the vanishing of the quasi-particle weight $Z_\sigma$. 
For momenta off the FS of species $\sigma$  but such that $\kLb - \sQfflo \leq \, k\, \leq  \kLb + \sQfflo$, we have found instead a marginal FL behavior for the imaginary part of the retarded self-energy

In addition, we have analytically derived the low energy  expression for the bosonic and fermionic propagators at criticality, from which we have extracted the values of the bosonic and fermionic dynamical critical exponents $z_b=z_f=3$ and the corresponding 
bosonic and fermionic damping terms $\sim \Om/q$ and $~\om^{2/3}$.

With respect to the fully self-consistent $t$-matrix approach, the present MSCT approximation has the advantage of allowing for more analytical control of the theory.
Leveraging on this advantage, in 3D we have derived an analytical expression for the imaginary part of the retarded self-energy at the FFLO transition at low frequencies. In this way, we have recovered the square root dependence on the frequency observed only numerically in the fully self-consistent study of Ref.~\cite{Pini2023}, and obtained the dynamical critical exponents $z_f=2 \,(=z_b)$.

For the 3D case on a lattice, we also evaluated the $2n$-point vertex corrections at all orders $n > 1$ of  the bosonic effective action for the FFLO pairing instability \cite{Pistolesi-1996, Strinati2000}. Using standard scaling arguments, we established that these corrections scale to zero in the long-wavelength and low-energy limit, having no impact on the critical behavior. Consequently, we place the 3D FFLO quantum-phase transition on a lattice in the mean-field universality class, in analogy  with the 3D itinerant antiferromagnet in heavy-fermion systems \cite{Steglich-2006}.

We have found that the origin of the differences  between 2D and 3D in the critical behavior of the bosonic and fermionic low energy excitations is in the geometry of the FFLO ordering momentum $\vQfflo$  with respect to the  momenta $\vkLup$, $\vkLdn$ at the points where the Fermi spheres match. In 2D, $\sQfflo=\kLup-\kLdn$ and the three momenta are collinear. As a result, the two FS's are tangent at the matching point and the local geometry can be approximated by a parabola. In 3D, $\sQfflo>\kLup-\kLdn$, the three momenta are not collinear and the two FS's intersect each other. These different geometries lead to different low-energy fermionic propagators to be considered in the convolution determining the bosonic propagator, and thus to a different form of the damping of critical fluctuations.

As a perspective, the implementation of the fully self-consistent $t$-matrix approach in 2D would be an important achievement. While the critical exponents and low energy propagators are expected to recover the results of the one-loop renormalization group calculation of Ref.~\cite{Pimenov_2018}, the implementation of the self-consistent $t$-matrix approach would allow for the calculation of non-universal quantities, such as the phase-diagram, thermodynamic variables, momentum distribution functions, and spectral weight functions in a full frequency range.     

Finally, we discuss how the predicted critical behavior could be probed in ultracold-atom experiments. Although the present calculations are performed at zero temperature, their low-energy predictions could be tested at finite but sufficiently low temperatures.
On the one hand, at finite temperature, the quantum critical behavior extends over a finite region around the critical polarization $p-p_c \sim k_BT/\eF$~\cite{Sachdev2011}. On the other hand, the temperature should be lower than the probed frequencies. 

The fermionic sector could be investigated using momentum-resolved radio-frequency spectroscopy, which provides access to the occupied single-particle spectral function  \cite{Jin-2015}. In the above temperature and polarization range, the measured line shapes could be used to test the predicted non-Fermi-liquid behavior
$
-\IM\Sigma^{\rm R}_\sigma(k_{{\rm F}\sigma},\bar{\omega})
\propto |\bar{\omega}|^{2/3},
$
in contrast with the Fermi-liquid dependence proportional to $\bar{\omega}^{2}$ away from criticality. Although both the linewidth and the excitation energy vanish as $\bar{\omega}\rightarrow 0$, their ratio diverges, signaling the absence of a long-lived quasiparticle. Fits to the coherent spectral feature extrapolated toward $T=0$ could be used to follow the predicted suppression $Z_\sigma\propto(p-p_c)^{1/2}$.

The same momentum-resolved spectra could be used to test the fermionic dynamical exponent $z_f$.
In the quantum critical region, the single-particle spectral function at low frequencies $\bar{\omega}$  becomes a universal function of $\omb / k^z$, where $k$ is the momentum distance from the Fermi surface, as detailed in Eq. \eqref{eq:Senthil}. The dynamical exponent $z_f$ could be extracted  as a fitting parameter by demonstrating the collapse of spectral functions measured at different momenta $k$ into a single universal curve when plotted against $\omb / k^{z_f}$.

Probing the bosonic sector requires access to pair correlations. 
Static pair correlations, measurable through atom-resolved microscopy \cite{Yao-2025,Feng-2026} or pair projection followed by momentum-space imaging \cite{Regal-2004,Ries-2015,Dyke-2021}, should develop a peak at $|\vQ|=Q_{\mathrm{FF}}$. The  width of this peak would allow the extraction of the correlation length and could therefore be used to test the exponent $\nu$. 

More directly, momentum-resolved Raman spectroscopy \cite{Wang-2012,Fu-2012,Shkedrov-2020} has recently been proposed as a probe of molecular spectral functions and unconventional pairing \cite{Diessel-2024} and could thus provide a route to accessing the momentum- and frequency-dependent pair spectral weight.
Similarly to the fermionic case, a scaling collapse of the molecular spectral functions measured at different momenta would provide a direct test of the bosonic dynamical exponent $z_b$.

Data for reproducing the figures are available online \cite{DatasetZenodo}.

\begin{acknowledgments}
We thank Davide Sottocorno for useful discussions and for checking parts of the analytic calculations.

L.P. and P.P. acknowledge financial support from the Italian Ministry of University and Research (MUR) under project PRIN2022, Contract No. 2022523NA7. P.P. also
acknowledges financial support from the European Union - Next Generation EU through MUR
projects PE0000023-NQSTI (Italy). 
\end{acknowledgments}

\appendix

\section{Pair susceptibility, generalized Thouless criterion and correlation length exponent $\nu$}
\label{sec:Thoulcrit}

    The pair-propagator\til\eqref{equ:gammamats} can be computed analytically following Ref.\til\cite{Pisani2025}, which employs the  NSCT approximation.  Its expression within the MSCT approximation can be obtained straightforwardly via the procedure illustrated in the second section of this appendix.

\subsection{ $\Gamma_0(\vQ,z)$ in the NSCT approximation}

Considering the extension of  $\Gamma_0(\vQ,z)$ to the upper complex plane ($\IM z >0$), one obtains the   expression
\begin{widetext}
\begin{equation}
\label{equ:gamma}
\Gamma_0(\vQ,z)^{-1} = -\frac{m}{4\pi}\left[ \log\left( \frac{A_0}{\be} \right)+\Theta(\mu_\uparrow) \log \left(\frac{A_\uparrow}{A_0}\right)  + \Theta(\mu_\downarrow) \log\left(\frac{A_\downarrow}{A_0}\right) -  \left[ \Theta(\mu_\uparrow) +\Theta(\mu_\downarrow) \right] \,i\,\pi\, \text{sgn}(\IM z)  \right], 
\end{equation}
where the principal branch of the complex $\log$ function (with the cut along the negative real axis) is meant in the above equation 
\begin{align}
    A_0&=\frac{Q^2}{4m} -2\mu-z, \label{equ:A0}\\
    A_\uparrow&=+h+\frac{z}{2}+\text{sgn}\left(+h-\frac{Q^2}{4m}+\frac{\RE z}{2} \right)\sqrt{ H_Q +z\left(h-\frac{Q^2}{4m}\right)+\frac{z^2}{4}},  \label{equ:Aup}  \\
    A_\downarrow&=-h+\frac{z}{2}+\text{sgn}\left(-h-\frac{Q^2}{4m}+\frac{\RE z}{2} \right)\sqrt{H_Q-z\left(h+\frac{Q^2}{4m}\right)+\frac{z^2}{4}},  
    \label{equ:Adn}
\end{align}
\end{widetext}
and 
\begin{equation}    
H_Q= \begin{cases}
    \left(\frac{Q^2-\bar{Q}^2}{4m}\right) \left(\frac{Q^2-{Q^*}^2}{4m}\right) & \text{if $\mu_\uparrow, \mu_\downarrow >0 $}, \\
    h^2+\frac{Q^2}{2m} \left(\frac{Q^2}{8m}-\mu\right) & \text{otherwise},
\end{cases}
\end{equation}
with $ \mu  \equiv(\mu_\uparrow+\mu_\downarrow)/2$ and $  h \equiv(\mu_\uparrow-\mu_\downarrow)/2. $ In the coupling regime where $\mu_\uparrow, \mu_\downarrow >0 $, it is convenient to introduce the notation
$
\Qfflo \equiv k_{\mu_\uparrow}-k_{\mu_\downarrow}$ and $  \bar{Q} \equiv k_{\mu_\uparrow}+k_{\mu_\downarrow}$, 
with $ k_{\mu_\sigma}=\sqrt{2m\mu_\sigma}$.
\begin{figure}[b]
    \centering
    \includegraphics[width=0.95\linewidth]{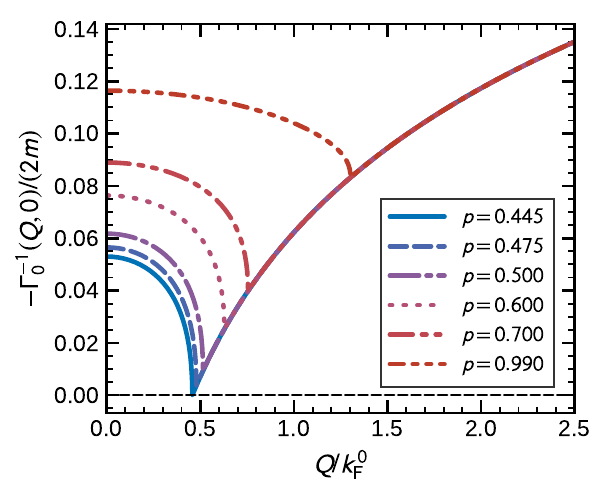}
    \caption{Inverse static pair susceptibility $-\Gamma_0(Q, i\Omega=0)^{-1}/2m$ as a function of $Q$ for several choices of the polarization $p$ at fixed coupling $\gtwoD=-1.48$ (corresponding to $\varepsilon_0/\eF=0.105$) within the MSCT approach.  }
    \label{fig:Thoulcr}
\end{figure}

We note that, in the limit of interest $z \rightarrow 0$ for the generalized Thouless criterion \eqref{equ:Thoulcr},  the function in  Eq.\til\eqref{equ:gamma} is not differentiable in $Q=\Qfflo$ and has the form  
\begin{equation}
-\Gamma_0(\vQ,0)^{-1} =  
\frac{m}{4\pi}
    \begin{cases}
      \ln\frac{Q^2}{4m\be}, &  Q \geq \Qfflo, \\
      \ln\frac{(h+\sqrt{H_Q})^2}{\be \left( 2\mu -\frac{Q^2}{4m} \right)} & Q \leq \Qfflo.
    \end{cases}
    \label{equ:gammaThoul}
\end{equation}
(Analogous expressions were found in  Ref.\til\cite{Sheehy2015} but with a different normalization, $m=1$).

Recalling that $-\Gamma_0(\vQ,0)^{-1}$ is the inverse of the static pair susceptibility, we observe from Eq.~\eqref{equ:gammaThoul} that for any value of $Q$ this quantity is positive, and for $Q=\Qfflo$ it has a minimum at which the left and right derivatives are not equal (see Fig.\til\ref{fig:Thoulcr}).
At this momentum, the susceptibility has a maximum, which reflects the enhanced tendency in the system to form pairs with finite momentum $\sQfflo \equiv \Qfflo = k_{\mu_\uparrow}-k_{\mu_\downarrow}$.
At the critical polarization, the pair susceptibility diverges and consequently the minimum of its inverse reaches zero and the generalized Thouless criterion is satisfied. Analytic expressions of the momentum $\sQfflo$ at criticality and of the critical pseudo-magnetic field $h_c$ are obtained by setting the arguments of both logarithms in \eqref{equ:gammaThoul} equal to one,
\begin{equation}
 \sQfflo = \sqrt{4m\be}, \quad  \quad h_c\equiv\sqrt{\be(2\mu-\be)}.
\label{equ:qchc}
\end{equation}

Within the NSCT approximation,  $\mu_\downarrow$ becomes negative along the second branch of Fig. \ref{fig:phasediag} and in this case the minimum of the inverse pair susceptibility is found at $Q=0$, 
\begin{equation}
-\Gamma_0(0,0)^{-1} =  
\frac{m}{4\pi} \ln{ \frac{2h}{\be} }. 
\end{equation}
Therefore, one obtains the critical magnetic field $h_c=\be/2$ with critical pair momentum $Q=0$, which identifies a transition to a spin polarized and spatially uniform superfluid (Sarma state) \cite{Sarma1963}.

\subsection{ $\Gamma_0(\vQ,z)$ in the MSCT approximation and correlation length exponent $\nu$}
\label{sec:appA2}

Within the MSCT approximation adopted in this work, the expressions in \eqref{equ:gamma} and in \eqref{equ:gammaThoul} should be rewritten by mapping
\begin{align}
    \mu_\sigma \rightarrow \muL, & \qquad
    k_{\mu_\sigma} \rightarrow \kL, 
    \label{equ:replmsct}
\end{align}
and thus
\begin{equation}
    \quad \mu \rightarrow \tilde\mu \equiv(\tilde\mu_\uparrow+\tilde\mu_\downarrow)/2, \quad h \rightarrow \tilde h \equiv(\tilde\mu_\uparrow-\tilde\mu_\downarrow)/2.  
    \label{equ:qffMSCT}    
\end{equation}
At criticality, one then has $ \sQfflo = \kLup-\kLdn = \sqrt{4m\be}$ and $\tilde h_c\equiv\sqrt{\be(2\tilde \mu-\be)}$.
The physical pseudo-magnetic field $h_c$ at criticality is instead defined in terms of the thermodynamic chemical potentials
\begin{equation}
    h \equiv\frac{\mu_\uparrow-\mu_\downarrow}{2}= \tilde h - \frac{\tilde\Sigma^0_\uparrow-\tilde\Sigma^0_\downarrow}{2},
\end{equation}
with $\tilde\Sigma^0_\sigma$ given in Eq.~\eqref{equ:s0til}.

Figure \ref{fig:Thoulcr} reports the behavior of the inverse pair susceptibility $-\Gamma_0(Q,  0)^{-1}/(2m)$ as a function of $Q$ for different values of the polarization $p$, for a representative choice of the coupling value 
$\gtwoD=-1.48$. The position of the cusp is traced  by the quantity $\sQfflo=\kLup-\kLdn$, which takes the value $\sQfflo=\sqrt{4m\be}$ when $p=p_c$.

For momenta close to the pairing momentum $\sQfflo$ and fields $h \simeq h_c$, Eq.~\eqref{equ:gammaThoul} acquires the form
\begin{align}
&-\Gamma_0(\vQ,0)^{-1} =  
\frac{m}{2\pi}\nonumber\\
   &\times \begin{cases}
     \frac{h-h_c}{ \tilde h_c}  +
    \frac{Q-\sQfflo}{\sQfflo}, &  Q \geq \sQfflo, \\
     \frac{h-h_c}{\tilde h_c}
     + \frac{1}{ \tilde h_c} \sqrt{ \frac{\kLdn}{m} \frac{\kLup^2}{2m} (\sQfflo- Q)  } & Q \leq \sQfflo,
    \end{cases}
    \label{equ:gammaThoulOZ}
\end{align}
where the mass term  $(h-h_c)/\tilde{h}_c $ sets the height of the cusp in Fig.~\ref{fig:Thoulcr}.

We are now in a position to obtain the critical exponent $\nu$ of the correlation length from Eq.~\eqref{equ:gammaThoulOZ}. 
At zero temperature, statics and dynamics cannot be treated separately \cite{Sondhi1997} and at criticality bosonic and fermionic correlations must be dealt with on the same footing. In the evaluation of the fermionic self-energy in appendix \ref{app:imsigretonoffFS} below, one sees from Fig. \ref{fig:hotspot} that the dominant bosonic contribution comes from fluctuations with momentum $\vQ$ that is near the intersection point between the red solid circle and the inner ring of the gray annulus, within the annulus. At this point the bosonic propagator diverges and this divergence is approached under the constraint $Q>\sQfflo$.
As a result, this selects a specific form for the static pair susceptibility, which is the one appearing in the first line of Eq.~\eqref{equ:gammaThoulOZ}.

Introducing, as in sec.~\ref{sec:dynexpB}, the radial ($q_r$) and transverse ($q_t$) components of the fluctuation momentum $\vq\equiv \vQ-\vQfflo$,  the deviation $Q-\sQfflo$ in Eq.~\eqref{equ:gammaThoulOZ} can be parametrized as follows,
\begin{equation}
    \lvert  \vQfflo + \vq \rvert - \sQfflo  \approx \left( \frac{q_t^2}{2 \sQfflo}
+ q_r \right),
\label{equ:qqfflo}
\end{equation}
having neglected the subleading term $q_r^2$ with respect to $q_r$.
The bosonic momentum $\vq+\vQfflo$ connects two fermionic momenta near the  matching point of the two FS's in Fig.~\ref{fig:FSgeom}(a). As the two FS's around this point extend primarily along the transverse direction (with respect to $\vQfflo$ which is along $x$), fluctuations possess a predominantly tangential component \cite{Sachdev2011,Lee-2008,Metlitski-2010,Senthil2015}, and therefore $q_t \gg q_r$. Neglecting the radial component of fluctuations in \eqref{equ:qqfflo},
the inverse static pair susceptibility \eqref{equ:gammaThoulOZ} acquires the familiar Ornstein-Zernicke \cite{Fisher1967} form  ($ [h-h_c]/\tilde{h}_c + q_t^2/2\sQfflosq$), from which
we extract the correlation length behavior $\xi \sim (h-h_c)^{-\nu} $ yielding $\nu=1/2$.

\section{Pair Spectral Weight Function} \label{sec:ImGammaRet} 
The pair spectral weight function $\IM \Gamma_0^{\rm R}(\vQ, \Omb)$ is obtained by taking the analytic continuation $z \to \Omb+i\,0^+$ in Eq.\til\eqref{equ:gamma}.

To gain insight into the origins of the various energy bands appearing in the intensity plot of Fig.\til\ref{fig:GammaPols} for $\IM \Gamma_0^{\rm R}(\vQ, \Omb)$, we perform the analytic continuation directly in Eq.\til\eqref{equ:gammamats}. Within the MSCT approximation, we obtain
\begin{widetext}
\begin{equation}
\IM[\Gamma_0(\vQ, i\Omega \to \Omb+ i\, 0^+)^{-1}] =  
     \int \frac{d\mathbf{k}}{(2\pi)^2} 
   \left[1 - \Theta(-\xit_{\vQ/2-\vk}^{\uparrow}) - \Theta(-\xit_{\vQ/2+\vk}^{\downarrow})\right] \;(-\pi)\; \delta\left(\xit_{\vQ/2-\vk}^{\uparrow} + \xit_{\vQ/2+\vk}^{\downarrow} - \Omb\right),
\label{eq:imgGamma}
\end{equation}
\end{widetext}
where  $\Theta$ is the Heaviside step function.
The momentum integral above splits into three distinct contributions corresponding to the three terms in the square brackets. The first term matches the $t$-matrix in vacuum while the remaining terms
account for the Pauli blocking of the  majority $\uparrow$ and minority $\downarrow$ species, respectively.
The first integral defines the overall continuum threshold for the two-body problem in vacuum, $\Omb_{\rm th}(Q)\equiv Q^2/4m-\mut_\uparrow-\mut_\downarrow$. For frequencies $\Omb< \Omb_{\rm th}(Q)$ one has $\IM[ \Gamma_0^{\rm R}(\vQ, \Omb)^{-1}]=0$. 

The second and third integrals are formally equivalent. Their $\Theta$ functions are non-zero when
\begin{equation}
\label{equ:costheta}
    -1 \leq  \frac{\mut_\uparrow-\mut_\downarrow-\Omb}{\sqrt{\frac{Q^2}{m}(\Omb-\Omb_{\rm th}(Q))}} \leq +1,
\end{equation}
which in turn becomes   
$
\Omb^-_\sigma(Q) \leq \Omb \leq \Omb^+_\sigma(Q),
$
with 
\begin{align}
    \Omb^\pm_\sigma(Q)& =\frac{(Q\pm\kL)^2}{2m}-\frac{\kLb^2}{2m},
\end{align}
where we have used $\mut_{\sigma}=\kL^2/2m$. When one $\Theta$ function is equal to one (and the other is zero), it cancels the 1 within the square bracket in Eq.~\eqref{eq:imgGamma}. This yields the regions labeled I and II in Fig.\til\ref{fig:GammaPols} where $\IM[ \Gamma_0^{\rm R}(\vQ, \Omb)^{-1}]=0$. In these regions, as in the region below $\Omb_{\rm th}(Q)$, the continuum of the pair spectral weight function identically vanishes. When the threshold frequencies are inside the continuum of  $\IM[ \Gamma_0^{\rm R}(\vQ, \Omb)$, they identify the position of cusps in $\IM[ \Gamma_0^{\rm R}(\vQ, \Omb)$. This information is used when performing numerical integrals such as in Eq.~\eqref{equ:sigfb}.

\begin{widetext}

\section{Low-energy limit of the critical fluctuation propagator}
\label{sec:gammalowE}

Here we analyze in detail the behavior of the fluctuation propagator, given in general by expression \eqref{equ:gamma}, when approaching criticality. We will work within the MSCT approximation (that is, we will apply the replacements \eqref{equ:replmsct} in Eq.~\eqref{equ:gamma}.
We are interested in momenta close to the critical pair momentum $Q \to  \sQfflo$ and in the low energy limit $z \rightarrow 0 + i \, 0^+$.
We point out that we will not make any assumption on the strength of the interaction or the amount of polarization, thus generalizing the analysis of \cite{Piazza2016}.

The coefficients $A_\uparrow, A_\downarrow$ in Eqs.\til\eqref{equ:Aup} and \eqref{equ:Adn} can be recast as follows,
\begin{equation}
A_\sigma
  = \zeta_\sigma\frac{\sQfflo \bar{Q}}{4m}
     + \frac{z}{2}
   +  \zeta_\sigma\mathrm{sgn}\left(
       \frac{\sQfflo \bar{Q} - Q^2}{4m}
       + \frac{\RE z}{2}
     \right)
   \sqrt{
       \frac{(Q^2 - \bar{Q}^2)}{4m}\frac{(Q^2 - \sQfflosq)}{4m}
       - z\frac{Q^2 -  \zeta_\sigma\sQfflo\bar{Q}}{4m}
       + \frac{z^2}{4}
     } .
\end{equation}
where $\zeta_\sigma=1$ or $-1$ if $\sigma=\uparrow$ or $\downarrow$, respectively.
For small $z$ and $Q \to \sQfflo$, it is easy to check that  the $\sgn$ functions are equal to 1 and, neglecting higher order terms in $z$ and $|Q-Q_{\rm FF}|$  
\begin{equation}
    A_\sigma
= \zeta_\sigma\frac{\sQfflo \bar{Q}}{4m} + \frac{z}{2}
     + \zeta_\sigma\sqrt{R_\sigma} 
\end{equation}
where 
\begin{equation}
R_\sigma
  = \frac{k_{{\rm F}\bar{\sigma}} \sQfflo}{2m} \left[- \frac{k_{{\rm F}\sigma}}{m} (Q - \sQfflo) +  \zeta_\sigma z \right] \label{eq:rup2} 
\end{equation}
while the coefficient $A_0$ in Eq.~\eqref{equ:A0} becomes
$
A_0 = 
           - \bar{Q}^2/4m - z .
$
By inserting these expressions  in Eq.~\eqref{equ:gamma}, and neglecting higher order terms, one obtains 
\begin{align}
 \Gamma_0(\vQ,z)^{-1}
  \simeq -\frac{m}{4\pi} \Biggl[
      -\ln\left( \frac{\bar{Q}^2}{4m\varepsilon_0} \right)
        + \ln\left(
          \frac{\sQfflo\bar{Q}}{4m\varepsilon_0}
          + \frac{\sqrt{R_\uparrow}}{\varepsilon_0}
        \right)
        + \ln\left(
          \frac{\sQfflo\bar{Q}}{4m\varepsilon_0}
          + \frac{\sqrt{R_\downarrow}}{\varepsilon_0}
        \right)
    \Biggr].
\end{align}    
where we have used $\Theta(\bar{\mu}_\sigma)=1$ and $\IM z > 0$ to cancel the last term in Eq.~\eqref{equ:gamma} with a corresponding term originating from the first $\log$ term in Eq.~\eqref{equ:gamma}. Using further that 
at criticality  $\sQfflo=\sqrt{4m\be}$, expanding the  last two log terms for $R_\sigma$ small, and neglecting higher order cross terms between $z$ and $(Q - \sQfflo)$, we obtain
\begin{equation}
\Gamma_0(\vQ,z)^{-1}
   = -\frac{m}{4\pi} \Bigl[
      a_{\downarrow}\sqrt{-e_Q^\uparrow +z }    + a_{\uparrow}\sqrt{-e_Q^\downarrow - z}   
   + b  \, v \, (Q - \sQfflo) 
    \Bigr],
\label{equ:gammaRle}
\end{equation}
where
\begin{equation}
e^\sigma_Q
  = \vL  (Q - \sQfflo), \quad
a_\sigma = \frac{\sqrt{\vL \sQfflo/2}}
         {v \, \sQfflo/2} ,
\end{equation}
with $v=(\vLup+\vLdn)/2$, $m\,\vL=\kL$ 
and $ b \equiv \frac{\vLup\, \vLdn}{v^2} \frac{2}{v\,\sQfflo}$.

The above expression for the critical pair fluctuation propagator generalizes the one obtained in \cite{Piazza2016} by relaxing the restrictive assumptions  therein of a weak interaction strength and small population imbalance.

Expression \eqref{equ:gammaRle} is also analogous to the critical fluctuation propagator for the incommensurate charge-density wave  order \cite{Holder2014}  and for the spin-density wave  order  \cite{Altshuler-1995}.
The analogy can be clarified by expanding expression \eqref{equ:gammaRle} for $|z| \ll |e_Q^\sigma|$
and $|e_Q^\sigma| \rightarrow 0$.
By setting $\vQ=\vQfflo + \vq$, where $\vq$ is small, one obtains
\begin{equation}
    e^{\sigma}_{Q} \approx
 \vL \left( \frac{q_t^2}{2 \sQfflo}
+ q_r \right),
\label{equ:eqfflo}
\end{equation}
where, as above, we have introduced the radial and transverse component of the fluctuation vector $\vq$ with respect to $\vQfflo$, and have  neglected a subleading term $\sim q_r^2$ with respect to $q_r$ in Eq.~\eqref{equ:eqfflo} \cite{Lee-2008,Senthil2015}. 
The approximate dispersion in \eqref{equ:eqfflo} is justified whenever the underlying FS has a non zero curvature. 

Inserting Eq. \eqref{equ:eqfflo} into Eq. \eqref{equ:gammaRle}
and expanding for $|z| \ll |e^{\sigma}_{Q}|$, we finally obtain 
\begin{equation}
\Gamma_0(\vQ,z)^{-1}=-\frac{m}{4\pi}
\begin{cases}
   -\,\frac{1}{\sQfflo \, \sqrt{\vLup \vLdn} } \,\frac{i \, z}{\left[ \left(\frac{q_t}{2\sQfflo}\right)^2+\frac{q_r}{2\sQfflo} \right]^{1/2} } \, + \frac{\vLup \vLdn}{v^2} \,   \left[ \left(\frac{q_t}{2\sQfflo}\right)^2+\frac{q_r}{2\sQfflo} \right],
    & \qquad e_Q^\sigma >0, \\
 \frac{1}{2\,m\, v \, \sqrt{\vLup\vLdn}} \, \frac{ z}{\left[ -\left(\frac{q_t}{2\sQfflo}\right)^2 - \frac{q_r}{2\sQfflo}  \right]^{1/2} } +  4   \frac{\sqrt{\vLup \vLdn}}{v} \, \left[-\left(\frac{q_t}{2\sQfflo}\right)^2-\frac{q_r}{2\sQfflo} \right]^{1/2},  &   \qquad e_Q^\sigma <0.
\label{equ:gammaRuniv2}
\end{cases}
\end{equation}
where we have used $\IM z > 0$.
A functionally analogous expression was obtained in the context of the incommensurate charge density wave order in \cite{Mandal2024}.
The above expression will be used in sec. \ref{sec:dynexpB} when discussing  the bosonic dynamical  exponent $z_b$.

\section{Small-frequency behavior of the imaginary part of the retarded self-energy at criticality in 2D}
\label{app:imsigretonoffFS}

In this appendix we obtain an analytical expression for the imaginary part of the retarded self-energy $\IM\Sigma_{\sigma}^\mathrm{R}(\vk, \omb)$ at criticality for small $\omb$ within the MSCT scheme.
In our treatment we closely follow Ref.~\cite{Holder2014}. 
We start from the general expression
\begin{equation}
\IM\Sigma_{\downarrow}^\mathrm{R}(\vk, \omb) = \int \frac{d^2k^{\prime}}{(2\pi)^2} \IM \Gamma_0^\mathrm{R}(\vk^{\prime}-\vk, \omb + \xit_{\vk^{\prime}}^{\uparrow})  \left[ \Theta(-\xit_{\vk^{\prime}}^{\uparrow})-\Theta(-\omb -\xit_{\vk^{\prime}}^{\uparrow})\right],
\label{equ:imsigret}
\end{equation}
where, for convenience, we have focused on the minority species and made the transformation $\vk\to-\vk$ in the corresponding expression  $\eqref{equ:imsigret0}$, taking advantage of its rotational invariance.

\subsection{Fermionic momentum on the FS}
\label{sec:imsigretonFS}
\begin{figure*}[t!]
\centering
\includegraphics[width=0.5\textwidth]{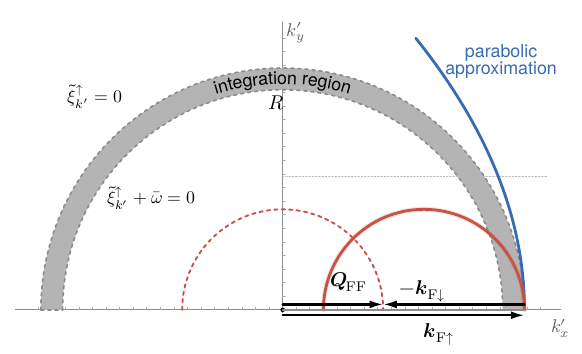}
     \caption{Gray annulus: 
     integration region $R$  for the momentum integral over $\vk'$ in Eq.~\eqref{equ:imsigret} at criticality, for $\vk=\vkLdn$  and $\omb>0$. (Only the upper half of the integration plane is shown for symmetry reasons.) Solid red circle: values of $\vk'$ for which the fluctuation propagator in Eq.~\eqref{equ:imsigret} is nearly divergent. When the solid red circle intersects the gray annulus the Thouless criterion is  exactly satisfied for given value of $\bar{\omega}$.
     The external momentum $\vkLdn$  is taken along the $x$ direction $k_x^\prime$ axis and is collinear with $\vkLup$ such that  $\kLup- \kLdn=\sQfflo$. Black dotted line: momentum cutoff $k_c$ along $k_y^\prime$. } 
     \label{fig:hotspot}
\end{figure*}
We first consider the case with $\vk$ on the FS: $\vk=\vkLdn$ such that  $|\vkLup|- |\vk|=\sQfflo\equiv\sqrt{4m\be}$. We start by examining the expression \eqref{equ:gammaRle} analytically continued to the real frequency axis ($z \to \bar{\Omega}+i0^+$)
\begin{align}
\Gamma^\mathrm{R}_0&(\vk^\prime - \vk , \omb + \xit_{ \vk^\prime}^{\uparrow})^{-1} = -\frac{m}{4 \pi} \Biggr[   a_{\uparrow}\sqrt{- e_{ |\vk^\prime- \vk|}^{\downarrow}- \omb - \xit_{ \vk^\prime}^{\uparrow} +i0^+}  + a_{\downarrow}\sqrt{- e_{|\vk^\prime- \vk|}^{\uparrow} + \omb + \xit_{ \vk^\prime}^{\uparrow} +i0^+} + b \, v \left( | \vk^\prime - \vk | - \sQfflo\right) \Biggr].
\label{equ:imgamretE}
\end{align}

Figure \ref{fig:hotspot} shows the integration plane $(k^\prime_x,k^\prime_y)$, with $\vk=\vkLdn$ setting the direction of the $k^\prime_x$ axis. For small frequency  $\omb$, the two $\Theta$ functions in Eq.~\eqref{equ:imgamretE} select the narrow annulus $ -\omb < \xit_{ \vk^\prime}^{\uparrow}<0$ around the FS $ |\vk| = \vkLup $, while the pair spectral weight function is strongly peaked close to the FFLO singularity of the pair propagator (see Fig.~\eqref{fig:GammaPols}) which in turns occurs when $| \vk^{\prime} - \vk| = \sQfflo$ (red solid circle).  
The dominant contribution to the convolution in Eq.~\eqref{equ:imsigret}  comes from the red circle within the annulus.

It is therefore convenient to shift $\vk^{\prime}\to \vk^{\prime} + \vkLup $ and 
expand the one- and two-particle dispersions $ \xit_{\vk^{\prime}}^{\uparrow}$  and $e^{\sigma}_{|\vk^\prime - \vk|}$ appearing in Eq.~\eqref{equ:imgamretE} about the new origin:
\begin{align}
\xit_{\vk^{\prime}}^{\uparrow} &\rightarrow \frac{|\vk^{\prime} + \vkLup|^2}{2m} - \frac{(\vkLup)^2}{2m} \approx \frac{{k^{\prime}_y}^2}{2m} + \vLup \,  k^{\prime}_x, \label{eq:newfs} \\
e^{\sigma}_{|\vk^\prime - \vk|}
&\rightarrow \vL  \left( \lvert \vk' + \vQfflo \rvert - \sQfflo \right) \approx
 \left(\frac{\vL}{\vff}\right) \, \frac{k_y'^2}{2 m} + \vL \,k_x',
\label{eq:neweQ}
\end{align}
with $\kL=m\,\vL$ and $\sQfflo=m \vff$. In Eq.~\eqref{eq:newfs} we have neglected the term ${k_x^{\prime}}^2/2m$, effectively approximating the FS to a parabola.
The validity of this approximation is limited to a region defined by $-k_c< k_y^{\prime} < k_c$ (blue line labeled ``parabolic approximation" in Fig.~\ref{fig:hotspot}) with $k_c$ an appropriate cut-off. (The value of $k_c$ is immaterial in the calculation that follows since $\omb$ can be chosen arbitrarily small.)
The parabolic approximation of the FS is familiar in the low energy effective theory of fermions coupled to a U(1) gauge boson in 2+1 dimensions \cite{Holder-2015,Lee-2009}. It is a direct consequence of the existence of fermionic soft modes in an isotropic system \cite{Sachdev2011,BKV-2005}. We note that such an approximation holds for the bosonic dispersion \eqref{eq:neweQ} as well (which in fact represents the order-parameter soft mode expected at the quantum phase transition), but with a curvature determined by the quantity  $\vLup / \vff$.

The self-energy convolution now acquires the form
\begin{align}
\IM\Sigma_{\downarrow}^\mathrm{R}(\vkLdn, \omb) \simeq & \int_R \frac{dk_x' \, dk_y'}{(2\pi)^2}\,
\frac{(-4\pi)}{m}\,
\IM \left\{\Bigg[
\; a_\uparrow \sqrt{- \frac{\vLdn}{\vff}
\left(\frac{k_y'^2}{2 m} + \vff \, k_x' \right)
- \left( \frac{{k^{\prime}_y}^2}{2m} + \vLup \,  k^{\prime}_x \right) - \omb - i0^+ } 
\nonumber \right.\\
&+\; \left. a_\downarrow \sqrt{- \frac{\vLup}{\vff}
\left( \frac{k_y'^2}{2m} + \vff \, k_x' \right)
+ \left( \frac{{k^{\prime}_y}^2}{2m} + \vLup \,  k^{\prime}_x \right) + \omb + i0^+ } 
+ b \,  \frac{v}{\vff} \left( \frac{k_y'^2}{2m} + \vff \, k_x'\right)\Bigg]^{-1}\right\},
\end{align}    
with $k_x^{\prime}$ and $k_y^{\prime}$ constrained within the region $R$ defined by $-2m\omb < {k_y^{\prime}}^2 + 2 m  \vLup k_x  < 0$, which reduces to $-k_c< k_y^{\prime} < k_c$ after the parabolic approximation. 

We now adopt the following change to dimensionless curvilinear coordinates $(\omt,\tilde{p})$ 
\begin{equation}
\begin{cases}
    \omt \, \omb = \frac{{k_y^{\prime}}^2}{2m} +   \vLup k_x^{\prime}\\
    \tilde{p} \, \frac{\vff}{v} \, \omb = \frac{{k_y^{\prime}}^2}{2m}+ \vff \,k_x^{\prime}.
\end{cases}    
\end{equation}
and modify the integration extrema accordingly. 
The integral acquires the form
\begin{align}
\IM\,\Sigma_\downarrow^\mathrm{R}(\vkLdn,\omb) = &
-4\, \frac{\omb}{mv}
\int_{-1}^{0} \frac{d\tilde{\omega}}{2\pi} 
\int_{ \frac{v}{\vLup} \omt }^{+\infty} d\tilde{p}
\; \; \sqrt{\frac{\frac{mv\vff}{2\vLup \vLdn }}{\tilde{p} - \frac{v}{\vLup} \omt } } \nonumber \\
& \times \IM \left\{\Bigg[   a_\uparrow \sqrt{
- \frac{\vLdn}{v}  \tilde{p} -\tilde{\omega}-1-i0^{+}} + a_\downarrow \sqrt{ - \frac{\vLup}{v}  \tilde{p} +\tilde{\omega}+1+i0^{+}} +
 b \, \sqrt{\omb} \;\tilde{p}
\Bigg]^{-1}\right\},
\label{integral}
\end{align}
where the square root term is the Jacobian of the transformation (assuming $ \vLup > \vLdn $), the upper bound of the inner integral has been extended to infinity as $\omb \to 0$, and we have taken into account the lower half of the integration region multiplicating the result of the integral by two.

The integral over $\tilde{p}$ in Eq.~\eqref{integral} is ultraviolet divergent in the limit $\omb \to 0$, with the dominant contribution to the integral 
coming from $\tilde{p} > \frac{1}{\sqrt{\omb}} \gg \tilde{\omega}$. We can then neglect the term proportional to $\tilde{\omega}$ in the Jacobian, approximate 
\begin{equation}
\label{eq:aexp}
  a_\uparrow \sqrt{
- \frac{\vLdn}{v}  \tilde{p} -\tilde{\omega}-1-i0^{+}}
+ a_\downarrow \sqrt{ - \frac{\vLup}{v}  \tilde{p} +\tilde{\omega}+1+i0^{+}}
        \approx 
-\sqrt{\frac{ v^2}{\vLdn \vLup}\frac{2}{m v \vff}}\, \frac{(1+\tilde{\omega})}{\sqrt{\tilde{p}}}i
\end{equation}
at large $\tilde{p}$, and set the lower limit of the momentum integral to zero. In this way one gets
\begin{align}
\IM\,\Sigma_\downarrow^\mathrm{R}(\vkLdn,\omb)
&= -\frac{4}{m \, v^2 \, b} \omb \int_{-1}^{0}  d\tilde{\omega} \,
   \int_{0}^{+\infty} \frac{d\tilde{p}}{2\pi}\,
   \IM\!\left\{\left[ -\frac{v^2}{\vLup \vLdn}\, (1+\tilde{\omega}) i +  \, \sqrt{b \, \omb} \, \tilde{p}^{3/2}
   \right]^{-1} \right\}.
\end{align}
The integrals over $\tilde{p}$ and $\tilde{\omega}$ are elementary and  eventually yield
\begin{align}
\IM\,\Sigma_\downarrow^\mathrm{R}(\vkLdn,\omb)
&=-\frac{1}{\sqrt{3}\,2^{4/3}}\, \frac{2m}{\kLdn \, \kLup}
\left(\frac{\kLupsq-\kLdnsq}{2m}\right)^{4/3} \, \omb^{2/3},
\qquad (\omb \to 0^+ ; \, p=p_{\rm c}).
\label{equ:imsigret23}
\end{align}
The above derivation can be repeated for $\omb <0$. In addition, by similar arguments one can show that the same expression \eqref{equ:imsigret23fin} also holds for the majority species, eventually yielding 
\begin{align}
\IM\,\Sigma_\sigma^{\rm R}(\vkLdn,\omb)&=-\frac{1}{\sqrt{3}\,2^{4/3}} \frac{2m}{\kLdn \, \kLup}
\left(\frac{\kLupsq-\kLdnsq}{2m}\right)^{4/3} \hspace{-0.3cm} |\omb|^{2/3}, \qquad (\omb \to 0 ; \, p=p_{\rm c}).
\label{equ:imsigret23fin}
\end{align}

\subsection{Fermionic momentum outside the FS}
\label{sec:imsigretoffFS}
\begin{figure*}[t!]
\centering
\includegraphics[width=0.5\textwidth]{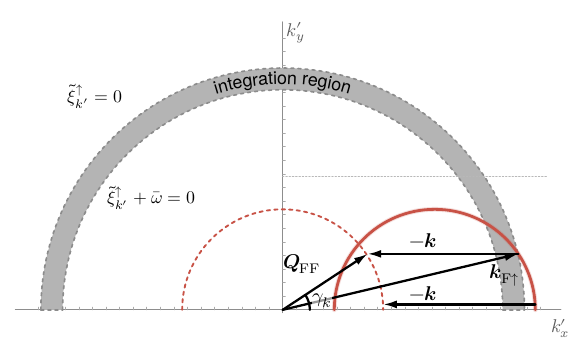}
\caption{Same as in Fig.\til\ref{fig:hotspot}, but for $\kLdn<k<2\kLup-\kLdn$ such that the pairing momentum $\vQfflo$ and the external momentum $\vk$  form an angle $\gamma_k\neq 0, \pi$.}
     \label{fig:hotspotnoncoll}
\end{figure*}

Focusing again on the minority species, we now consider the case $\kLdn < k < 2\kLup-\kdn$. In this case, a solution of the equation $|\vk'-\vk|=Q_{\rm FF}$ for the integration variable $\vk'$ exists when the (arbitrary) direction of $\vQ_{\rm FF}$ is chosen such that the angle $\gamma_k$ between $\vQfflo$ and $\vk$ satisfies Eq.~\eqref{equ:condfflo2}. As a result, $\gamma_k\neq 0, \pi$. 
The geometry then changes from the one in Fig.\til\ref{fig:hotspot} to that in Fig.\til\ref{fig:hotspotnoncoll}. 
One sees that the solid red circle and the FS (corresponding to the outer circle of the annulus) are no longer tangent, but they intercect each other.

We thus reconsider expressions  \eqref{eq:newfs} and \eqref{eq:neweQ} according to the geometry of  Fig.\til\ref{fig:hotspotnoncoll},
\begin{align}
\xit_{\vk^{\prime}}^{\uparrow} & \rightarrow \frac{|\vk^{\prime} + \vkLup|^2}{2m} - \frac{(\vkLup)^2}{2m} \approx  \vLupx \, k^{\prime}_x+\vLupy k^{\prime}_y , \label{eq:newfs2} \\
e^{\sigma}_{|\vk^\prime - \vk_\sigma|}
&\rightarrow \vL \left( \lvert \vk' + \vQfflo \rvert - \sQfflo \right) \approx  \vL \left( \cos \gamma_k \, k_x' 
+ \sin \gamma_k \, k_y' \right),
\label{eq:neweQ2}
\end{align}
whereby, owing to the non-collinearity between $\vk$, $\vQfflo$ and $\vkLup$, both linear components $k_x'$ and $k_y'$ are now present and one can therefore neglect the corresponding quadratic contributions ${k^{\prime}_x}^2$ and ${k^{\prime}_y}^2$.
The previous change to curvilinear coordinates is now a simple rotation of the $(k_x',k_y')$ plane according to the transformation
\begin{equation}
\begin{cases}
    \omt \, \omb  =\vLupx \, k^{\prime}_x+\vLupy k^{\prime}_y \\
    \tilde{p} \, \frac{\omb}{v} =  \left( \cos \gamma_k \, k_x'  + \sin \gamma_k \,  k_y' \right),
\end{cases}   
\end{equation}
which yields the following expression at $p=p_{\rm c}$:
\begin{align}
\IM\,\Sigma_\downarrow^\mathrm{R}(\vk,\omb)
= &-8\, \frac{\omb^{3/2}}{mv}
\int_{-1}^{0} \frac{d\tilde{\omega}}{2\pi} 
\int_{ \frac{v}{\vLup} \omt }^{+\infty} d\tilde{p}\;  \frac{m}{ |\vk| \, \sin \gamma_k } \nonumber \\
&\times \IM\left\{\Bigg[ 
  a_\uparrow \sqrt{
- \frac{\vLdn}{v}  \tilde{p} -\tilde{\omega}-1-i0^{+}}
 + a_\downarrow \sqrt{ - \frac{\vLup}{v}  \tilde{p} +\tilde{\omega}+1+i0^{+}} +
 b \, \sqrt{\omb} \;\tilde{p}
\Bigg]^{-1}\right\},
\end{align}
where the additional factor of 2 takes into account the presence of two matching points between the two FS rather than the single point of the collinear case, and the Jacobian of the transformation
($m/|\vk| \, \sin \gamma_k$) has been included.

\begin{figure*}[t!]
\centering
\includegraphics[width=0.90\linewidth]{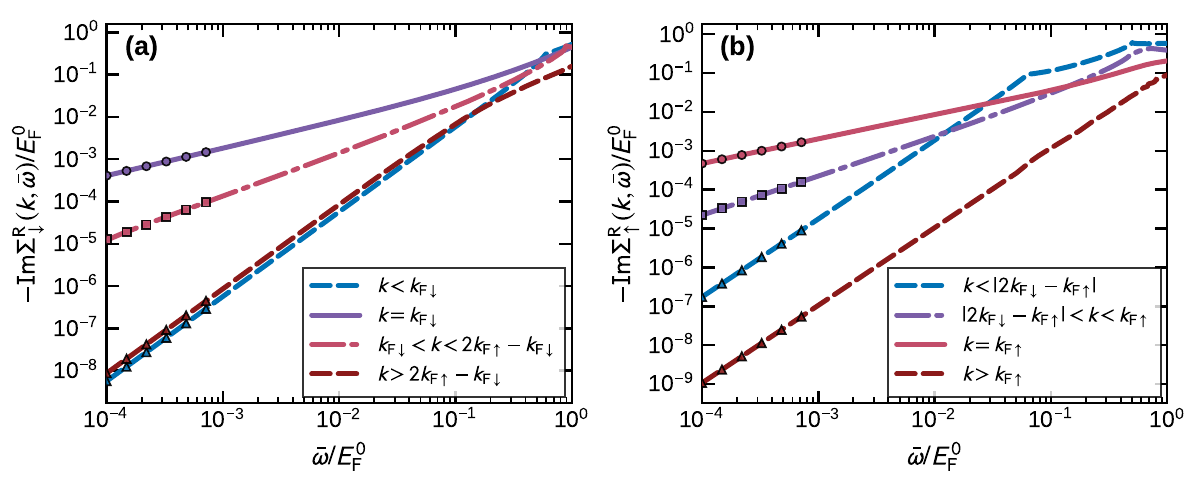}
     \caption{Imaginary part of the retarded self-energies as a function of frequency for (a) minority and (b) majority species at $\gtwoD=-1.48$ and $p=p_c$ for different values of $k$ (see legend). The symbols highlight the expected analytical low-frequency behavior for given $k$: $\propto \omb^{2/3}$ for $k=\kL$ (circles); $\propto \omb $ 
     for $|\kLb - \sQfflo|<  k  < \kLb + \sQfflo$ (squares)
     and $\propto \omb^{2}$ for (a) $k < \kLdn$ or $k > 2\kLup- \kLdn$; (b) $k < |2\kLdn -\kLup|$ or $k > \kLup $ (triangles).}
     \label{fig:xoverNFL-FL}
\end{figure*}

Applying the same expansion as in \eqref{eq:aexp}, one gets
\begin{align}
\IM\,\Sigma_\downarrow^\mathrm{R}(\vk,\omb)
&= -8 \,\frac{\omb^{3/2}}{mv} \int_{-1}^{0} d\tilde{\omega}  \,
   \int_{0}^{+\infty} \frac{d\tilde{p}}{2\pi} \; \frac{m}{ |\vk| \, \sin \gamma_k } \;
   \IM\!\left\{\left[ -\sqrt{\frac{ v^2}{\vLdn \vLup}\frac{2}{m v \vff}}\, \frac{(1+\tilde{\omega})}{\sqrt{\tilde{p}}}i
    +  b \, \sqrt{\omb}\;\tilde{p}
   \right]^{-1}\right\}.
\end{align}    
The two integrals are again elementary and yield
\begin{align}
\IM\,\Sigma_\downarrow^\mathrm{R}(\vk,\omb) = -\frac{2}{3}\, \frac{ v^2 \, m  \vff}{ \vLup \vLdn |\vk| \sin \gamma_k}\; \omb  
= -\frac{1}{6}\,\frac{(\kLupsq-\kLdnsq)^2}{|\vk \times \vQfflo| \, \kLup \kLdn } \; \omb,  \qquad (\omb \to 0^+ \, , \, \kLdn < k <  2\kLup -\kLdn\;  \, p=p_{\rm c}).
\end{align}    
As in the previous case, the above derivation is unchanged when $\omb \to -\omb$. 
In addition, the case of the majority species can be treated in the same way. In summary, one obtains 
\begin{align}
\IM\,\Sigma_\sigma^\mathrm{R}(\vk,\omb) 
= -\frac{1}{6}\,\frac{(\kLupsq-\kLdnsq)^2}{|\vk \times \vQfflo| \, \kLup \kLdn } \; |\omb|, 
\qquad (\omb \to 0  \, , \,  |\kLb - \sQfflo | < k <  \kLb + \sQfflo; \, p=p_{\rm c}). 
\label{equ:imsigretoffFS}
\end{align}
Finally, outside the above range of $k$, that is, for $k<|\kLb - \sQfflo |$ or $k > \kLb + \sQfflo$ it is easy to show that a standard Fermi liquid behavior is recovered, with $\IM\Sigma_\sigma^\mathrm{R}(\vk,\omb)\propto\bar{\omega}^2$ at small $\omb$.

Figure \ref{fig:xoverNFL-FL} compares the analytic results of the present appendix with a fully numerical calculation of $\IM \Sigma_\sigma^\mathrm{R}(\vk,\omb)$ for coupling strength  $\gtwoD=-1.498$  at $p=p_{\rm c}$.

\section{Imaginary part of the retarded self-energy at criticality in 3D}
\label{app:imsigret3D}

In this appendix we obtain an analytical expression for the imaginary part of the retarded self-energy at criticality in 3D in the MSCT approach  and in the limit of small frequency $\omb \to 0$.

We recall the expression of the fluctuation propagator in the imaginary frequency space in the low energy limit \cite{Perali2002,Pini2023}  
\begin{equation}
    \Gamma_0(\vQ,i\Omega)=\frac{1}{a+b\,(|\vQ|-\sQfflo)^2+(d_1+i\,d_2 \,\text{sgn}[\Omega])\,i\,\Omega},
    \label{equ:gammats3Dapp}
\end{equation}
where $a=0$ at criticality.
The imaginary part of the retarded self-energy is  given by
\begin{equation}
\IM\Sigma_{\sigma}^\mathrm{R}(\vk, \omb) = \int \frac{d^3Q}{(2\pi)^3} \IM \Gamma_0^\mathrm{R}(\vQ, \omb + \xit_{\vQ-\vk}^{\bar\sigma})\left[ \Theta(-\xit_{\vQ-\vk}^{\bar\sigma})-\Theta(-\omb -\xit_{\vQ-\vk}^{\bar\sigma})\right],
\label{equ:imsigret3D}
\end{equation}
where the retarded pair propagator is obtained by analytic continuation of Eq.\til\eqref{equ:gammats3Dapp}, replacing $i\Omega \to \Omb+i0^+$.
Given the linear relation between 
$\xit_{\vQ-\vk}^{\bar\sigma}$ and the cosine of the angle between $\vQ$ and $\vk$, the following change of  variable is adopted
\begin{equation}
    \cos\theta \rightarrow \xit \equiv  \xit_{\vQ-\vk}^{\bar\sigma}= \frac{Q^2+k^2-2Qk\cos\theta}{2m}-\tilde{\mu}_{\bar\sigma},
\end{equation}
thus yielding
\begin{equation}
\IM\Sigma_{\sigma}^\mathrm{R}(\vk, \omb) = \int_0^{+\infty} \frac{dQ}{(2\pi)}Q^2 \, \int_{\xit_-(Q)}^{\xit_+(Q)} \frac{d\xit}{(2\pi)}  \left( \frac{m}{Q k}\right) \frac{\left[-d_2 \,(\omb + \xit)\right] \; \left[ \Theta(-\xit)-\Theta(-\omb -\xit)\right]}{\left[ b\,(Q-\sQfflo)^2+d_1\,(\omb +\xit)\right]^2 +\left[d_2\,(\omb +\xit)\right]^2},   
\end{equation}
with $\xit_{\pm}(Q)=(Q\pm k)^2/2m-\tilde{\mu}_{\bar\sigma}$ and $\tilde{\mu}_\sigma=(\kL)^2/2m$ in the MSCT scheme.
The presence of the two $\Theta$ functions requires $-\omb < \xit < 0$ 
if $\omb>0$, or   $0 < \xit < -\omb$
if $\omb<0$, for the integrand to be non-vanishing. Considering that $\omb \to 0$ and neglecting subleading contributions to the integral, the boundaries of the integral over $\xit$ can be taken as determined from the above condition, while $|k-\kLb| < Q < k+\kLb$.
The integral over $\xit$ can be computed analytically and yields
\begin{equation}
\IM\Sigma_{\sigma}^\mathrm{R}(\vk, \omb) =
\int_{|k-\kLb|}^{k+\kLb} \frac{dQ}{(2\pi)^2}\, \frac{m Q}{k} \times
\begin{cases}
F\!\left(\dfrac{\omb}{(Q-\sQfflo)^2}\right), & \omb \geq 0, \\[2ex]
F\!\left(\dfrac{2\omb}{(Q-\sQfflo)^2}\right) - F\!\left(\dfrac{\omb}{(Q-\sQfflo)^2}\right), & \omb < 0.
\end{cases}
\label{equ:imsigret3DQ}
\end{equation}
where 
\begin{align}
F(x) =  \frac{1}{2  \left(d_1^2+d_2^2\right)} 
\left[ d_2 \log \left(\frac{ \left(d_1^2+d_2^2\right)}{b^2} x^2 +
\frac{2 d_1}{b} \, x +1\right) - 2 d_1 \arctan\left(\frac{\left(d_1^2+d_2^2\right)}{b \,d_2} \, x+\frac{d_1}{d_2}\right)+2 d_1 \arctan\left(\frac{d_1}{d_2}\right)\right].
\end{align}

The integrand is singular if $|\kLb - \sQfflo| < k < \kLb + \sQfflo$, provided that $\sQfflo$ is in the range of integration. We then assume $k$ to be in this range and change the integration variable to $t = (Q-\sQfflo)/\sqrt{|\omb|}$ 
\begin{equation}
\IM\Sigma_{\sigma}^\mathrm{R}(\vk, \omb)=
\sqrt{|\omb|} \, \int_{-\infty}^{+\infty} \frac{dt}{(2\pi)^2} \, \frac{m\,(\sQfflo+\sqrt{|\omb|} \, t)}{k} \,  \left[\sgn \omb \, F(\sgn \omb \, t^2) +\Theta(-\omb)F(-2 t^2)\right],
\label{equ:imsigret3Dt}
\end{equation}
where the lower and upper boundaries of the integration interval after the transformation take into account that $\omb \to 0$ and $k$ is in the above range.
The term proportional to $t$ does not contribute to the integral because it is odd in $t$. We thus obtain 
\begin{equation}
    \IM\Sigma_{\sigma}^\mathrm{R}(\vk, \omb) =\,\sQfflo \, \frac{ \sqrt{|\omb|} }{k}  \, I_\pm(b,d_1,d_2)
    \label{equ:imsigR3d}
\end{equation}
where $I_\pm$ identifies two functions of the parameters $b,d_1,d_2$ characterizing the critical fluctuation propagator for $\omb >0$ and $\omb<0$, respectively.

Therefore, the imaginary part of the retarded self-energy is a universal function of $\sqrt{|\omb|}/k$ whenever $|\kLb - \sQfflo| < k < \kLb + \sQfflo$. By taking $\vk$ on the FS and performing the analytic continuation to the positive imaginary frequency axis, this behavior leads to the low energy dressed fermionic propagator \eqref{equ:dressedG}, implying a fermionic dynamical exponent $z_f=2$.
In particular, the coefficients $B_\sigma$ and $C_\sigma$ of the self-energy on the imaginary frequency axis appearing in \eqref{equ:dressedG} can be determined following the analytic continuation procedure outlined in \cite{Pini2023},
from which we have explicitly
\begin{align}
    B_\sigma&=\frac{\sQfflo}{\sqrt{2} \, \kL} (I_+ - I_- ), \\
    C_\sigma&=\frac{\sQfflo}{\sqrt{2} \, \kL} (I_+ + I_- ).
\end{align}
In the complementary range of $k$, $ |k-\kLb| > \sQfflo$, the integrand in Eq.\til\eqref{equ:imsigret3DQ} remains finite and 
can be expanded in powers of $\omb$ yielding the Fermi liquid behavior $\IM\Sigma^{\rm R}(\vk,\omb)\propto \omb^2$.

\section{4-point vertex of the Hertz-Millis theory}
\label{app:4bosvert}

In this appendix we evaluate the lowest order vertex correction to the fluctuation propagator of the FFLO transition on a 3D lattice and show its irrelevance in the renormalization group flow associated  with the Hertz-Millis effective action in the pairing channel.
Within a functional integral formulation, the lowest order vertex correction to the pair propagator can be read from the $\phi^4$ term of the Hertz-Millis effective action for a 3D attractive Fermi gas. 
The form of this functional was  discussed in the context of the BCS-BEC crossover in \cite{Zwerger1992,Pistolesi-1996,Strinati2000} for the balanced system. In the present case, accounting for zero temperature and a population imbalance, it acquires the form
\begin{equation}
    S_\mathrm{eff}= v_0 \, \int  \frac{d^4Q}{(2\pi)^4} \; \Gamma_0^{-1}(Q) \;  |\Psi_{Q}|^2 + \frac{v_0^2}{4} \int  \prod_{i=1}^{4} \frac{d^4Q_i}{(2\pi)^4} \;
     u_2(Q_1,Q_2,Q_3,Q_4) \; \delta^{(4)}(Q_1+Q_2-Q_3-Q_4) \; \Psi^*_{Q_1}\,\Psi^*_{Q_2}\,\Psi_{Q_3}\,\Psi_{Q_4},
\label{equ:seff}
\end{equation}
where $Q_i\equiv(\vQ_i,i\Omega_i)$ are four-momenta, $\Psi_{Q_i}$ are the bosonic pairing fields, and
we have made  the dependence on the underlying bare fermionic coupling constant $v_0$ explicit. 

On a 3D lattice, the set of admissible ordering vectors is finite and dictated by the crystal symmetry, in contrast with the continuum, where $\vQfflo$ spans an entire spherical surface.
Moreover, the absence of continuous rotational symmetry changes the dispersion term of the fluctuation propagator $\Gamma_0^{-1}(Q)$ with respect to that in Eq.~\eqref{equ:gammats3D}, which becomes anisotropic. Nevertheless, the scaling exponents derived in Sec.~\ref{sec:boszvert} are unchanged in the discrete geometry, since the lattice only affects the non-universal coefficients of the dispersion term.

The 4-point vertex (depicted diagrammatically in Fig.\til\ref{fig:b4diag}) represents an effective boson-boson (or mode-mode) interaction \cite{Strinati2000}. It reads
\begin{equation}
u_2(Q_1,Q_3+Q_4-Q_1,Q_3,Q_4) = \int \frac{d^4k}{(2\pi)^4}\, \Gt_{0\uparrow}(-k)\, \Gt_{0\downarrow}(k+Q_1) \, \Gt_{0\uparrow}(-k-Q_1+Q_3) \, \Gt_{0\downarrow}(k+Q_4),
\label{equ:v4}
\end{equation}
with $k\equiv(\vk,\omega)$ and where the four-momentum conservation  $Q_1+Q_2=Q_3+Q_4$ has been used to eliminate $Q_2$. 

When discussing the corresponding problem in the particle-hole channel, Hertz assumes that the 4-point vertex is a well-behaved function of the four-momenta $Q_i$, and proceeds to evaluate it at $Q_i=0$ in the particle-hole channel (as relevant for an itinerant ferromagnet at the critical point),
thus finding a finite quantity \cite{Hertz-1974}. 
In the present case of an instability in the particle-particle channel with a non-zero ordering wave-vector,
it is straightforward to verify that the expression \eqref{equ:v4} at $\vQ_i=\vQfflo$ and zero frequencies is finite, apparently in full in analogy with the result of Hertz in the particle-hole channel. However, we will now show that the limit $\Omega_i \to 0$ for $\vQ_i=\vQfflo$ is actually singular, thus invalidating a treatment \emph{a la} Hertz.
Hertz's assumption of a well-behaved 4-point vertex was originally challenged by Chubukov in 2D {\em et al.} \cite{Chubukov-2003}  and subsequently in 3D \cite{Chubukov-2004-AFM}, demonstrating that this vertex depends singularly on the frequency-to-momentum ratio.

We thus proceed to evaluate the 4-point vertex
by keeping the four-momenta generic. One has
\begin{multline}    
    u_2(Q_1,Q_3+Q_4-Q_1,Q_3,Q_4) = \int \frac{d\omega}{2\pi}\frac{d^3k}{(2\pi)^3}\\ \times  \frac{1}{-i \omega-\tilde\xi^\uparrow_\vk}\, \frac{1}{i(\omega+\Omega_1)-\tilde\xi^\downarrow_{\vk+\vQ_1}} \; \frac{1}{-i(\omega+\Omega_1-\Omega_3)-\tilde\xi^\uparrow_{\vk+\vQ_1-\vQ_3}}  \, \frac{1}{i(\omega+\Omega_4)-\tilde\xi^\downarrow_{\vk+\vQ_4}}
\end{multline}
where we have used the form \eqref{equ:got3D} with linearized dispersions $\tilde\xi_\vk^\sigma=\vvL \cdot ( \vk-\vkL)$ for the bare-like propagators $\tilde{G}_0$ in Eq.~\eqref{equ:v4}.
We note that this expansion around a specific matching point $\vQfflo$ of the two FS's is justified only when the system selects a specific ordering wavevector out of a finite set of symmetry-related ones -- a condition that is met during a transition to the FF state on a lattice.

We shift  $\vk \to \vk+\vkLup$ to obtain $\tilde\xi_\vk^\uparrow \to \epsilon_\vk^\uparrow=\vvLup \cdot \vk$ and $\tilde\xi_\vk^\downarrow \to \epsilon_\vk^\downarrow=\vvLdn \cdot (\vk+\vQfflo)$. Recalling that $\vvL$ is in the $xy$ plane, we then make a change of variables from $k_x$, $k_y$  to $\epsilon_\uparrow,\epsilon_\downarrow$, where $\epsilon_\sigma=\vvL\cdot \vk$,  with the  resulting jacobian $|\vvLup \times \vvLdn |^{-1}$.
Introducing the large momentum cutoff $\Lambda$  along $k_z$, the integral now reads
\begin{align}    
     &u_2(Q_1,Q_3+Q_4-Q_1,Q_3,Q_4)  = \frac{1}{|\vvLup \times \vvLdn |} \int \frac{d\omega} {2\pi} \,
    \int_{-\Lambda}^{\Lambda}\frac{dk_z}{2\pi}\;
    \int \frac{d\epsilon_\uparrow}{2\pi}\, \frac{1}{i \omega+\epsilon_\uparrow}\, \frac{1}{i(\omega+\Omega_1-\Omega_3)+\epsilon_\uparrow+\epsilon^\uparrow_{\vQ_1-\vQ_3}}  \notag \\
            & \qquad \qquad \qquad \qquad \qquad \qquad \qquad \qquad \qquad \qquad\times 
    \int \frac{d\epsilon_\downarrow}{2\pi}\,\frac{1} {i(\omega+\Omega_1)-\epsilon_\downarrow-\epsilon^\downarrow_{\vQ_1}}  \frac{1}{i(\omega+\Omega_4)- \epsilon_\downarrow-\epsilon^\downarrow_{\vQ_4}} \\
    & \qquad \qquad =  \frac{1}{|\vvLup \times \vvLdn |} \frac{\Lambda}{\pi} \, \int \frac{d\omega} {2\pi} 
    \frac{\sgn(\omega)-\sgn(\omega+\Omega_1-\Omega_3)}{i(\Omega_1-\Omega_3)+\vvLup \cdot (\vQ_1-\vQ_3)} \;\;  \frac{\sgn(\omega+\Omega_1)-\sgn(\omega+\Omega_4)}{i(\Omega_1-\Omega_4)-\vvLdn \cdot (\vQ_1-\vQ_4)}.
\end{align}
The integration over $\omega$ is elementary and yields 
\begin{align}
    u_2(Q_1,Q_3+Q_4-Q_1,Q_3,Q_4) 
    =\frac{\Lambda}{\pi^2} \,\frac{1}{|\vvLup \times \vvLdn |} 
    \, \frac{|\Omega_3|+|\Omega_4|-|\Omega_1|-|\Omega_3+\Omega_4-\Omega_1|}{[i(\Omega_1-\Omega_3)+\vvLup \cdot (\vQ_1-\vQ_3)]\,[i(\Omega_1-\Omega_4)-\vvLdn \cdot (\vQ_1-\vQ_4)]}
    \label{equ:u4}
\end{align}
\end{widetext}
The expression above shows that $u_2$ becomes singular in the limit of small frequencies when all momenta are equal, and in particular for $\vQ_i=\vQfflo$. In contrast, Hertz assumed the quartic coefficient to be well-behaved in the limit of zero momenta and frequencies and calculated it at exactly zero momenta and frequencies \cite{Hertz-1974}.

Therefore a direct inspection of each singular term of the expansion is required to verify its correct scaling behavior by the power counting technique.
From the non-singular (gaussian) term in Eq.~\eqref{equ:seff}, taking into account that  $[\Omega]=z_b$, we obtain that the fields $\Psi_Q$ scale as
\begin{equation}
\left[\Psi_Q^2\right]=-\{  \underbrace{3+z_b}_{\textstyle [d^3Q\,d\Omega]} +\underbrace{2}_{\textstyle [\Gamma_0^{-1}]}\} = -5- z_b,    
\end{equation}
where $[X]$ indicate the scaling dimension of $X$. This is in line with Hertz's arguments. 
On the contrary, the quartic term dictates a new scaling of the renormalized coupling constant $v_0^2 \to g_4$, namely
\begin{equation}
[g_4]=-\{ \underbrace{(3+z_b)\,3}_{\textstyle[(d^3Q\,d\Omega)^{4-1}]}+\underbrace{z_b-2}_{\textstyle[u_2]}+\underbrace{2\,(-5-z_b)}_{\textstyle[\Psi_Q^4]} \}= 3-2z_b.   
\label{equ:uscale}
\end{equation}
Here, $[u_2]=z_b-2$ is obtained from the numerator of \eqref{equ:u4} scaling as $z_b$, and from the scaling of the denominator of \eqref{equ:u4}, where $[\epsilon_\sigma]=1$ dominates over $[\Omega]=z_b > 1$.
In the present case $z_b=2$, therefore $[g_4]=-1 < 0$
and the 4-point coupling constant $g_4$ is  irrelevant. The same scaling of $g_4$ would be obtained within Hertz's theory, which predicts $[g_4]=4-d-z_b$ (as can be obtained from \eqref{equ:uscale} by removing the contribution of $[u_2]$ and taking $d=3$). However, as already pointed out above, the assumption of regularity of the vertex $u_2$ used in Hertz's theory  is not valid in the present case (as is also the case for the IAFM \cite{Chubukov-2004-AFM}).  
The coincidence of the two outcomes is purely accidental and arises because  the case $z_b=2$ is very particular: the present scaling $[g_4]=6-d-2z_b$ (extended to dimension $d$)  and Hertz's scaling $[g_4]=4-d-z_b$ exactly coincide in any dimension when $z_b=2$. The peculiarity of the case $z_b=2$ was originally noticed for the itinerant antiferromagnet in \cite{Chubukov-2004-AFM}.

\section{Analytic expression of the critical polarization in the weak coupling regime}
\label{sec:wclim}

In the weak-coupling limit ($g\to -\infty$, with $g\equiv \gtwoD)$, the pair propagator can be approximated to a constant $\Gamma^0(\vQ,i\Omega) \simeq 2\pi/(m\,g)$ \cite{Pieri2024}, yielding the self-energy 
\begin{equation}
\Sigma_{\sigma}(\vk,i\omega) \simeq \frac{2\pi}{m\,g} \, n^0_{\mu_{\bar \sigma}} \equiv \Sigma_\sigma^0, 
\end{equation}
with $n^0_{\mu_{\bar \sigma}} = m \mu_{\sigma}/(2\pi)$.
To leading order in $1/g$, one can approximate 
$
 n^0_{\mu_{\sigma}} \simeq m \,\eFs/(2 \pi) \equiv n_\sigma.
$
The number equation \eqref{equ:dens} then yields $n_{\sigma}=n^0_{\mu^\prime_\sigma}$
with $\mu^\prime_\sigma = \mu_{\sigma} - 2\pi \, n_{\sigma}/(m\,g)$.
One then obtains
\begin{equation}
 \mu_{\sigma} = \frac{2\pi}{m}n_{\sigma} + \frac{2\pi}{m\,g} n_{\bar \sigma},
\end{equation}
which can be inserted in the expression for the critical field 
$h_c$ in\til\eqref{equ:qchc} to obtain an equation that implicitly defines $p_c$  
\begin{equation}
    p_c\left(1-\frac{1}{g}\right)=\sqrt{2 \left( \frac{\be}{\eF}\right) \left( 1+ \frac{1}{g} \right) - \left( \frac{\be}{\eF}\right)^2}.
\end{equation}
Using $\be/\eF=2\,e^{2g}$, one obtains 
\begin{equation}    
p_c= \frac{2g e^g}{g-1} \sqrt{1 - e^{2g} + \frac{1}{g}} \simeq \left[ 2 + \frac{3}{g}\right] e^g,
\end{equation}
to leading order  in $1/g$.

\bibliography{biblio}
\end{document}